\documentclass[prx,reprint,twocolumn,amsmath,amssymb]{revtex4}
\def\pdftitle{Temperature dependence of energy barrier and switching field of
sub-micron magnetic islands with perpendicular anisotropy}
\def\authorname{Leon Abelmann}
\def\pdfsubject{}
\def\pdfkeywords{}
\def\pdfbackref{none}

\usepackage[pdftex,
        colorlinks=true,
        urlcolor=darkblue,               
        anchorcolor=darkblue,
        filecolor=green,                   
        linkcolor=darkblue,              
        menucolor=darkblue,
        citecolor=darkblue,
        pagebackref,
        backref={\pdfbackref},
        pdfpagemode={UseOutlines},
        bookmarks=true,
        bookmarksopen=true,
        pdftitle={\pdftitle},
        pdfauthor={\authorname}, 
        pdfsubject={\pdfsubject},
        pdfkeywords={\pdfkeywords}
        ]{hyperref}

\usepackage{graphicx}
\DeclareGraphicsExtensions{.pdf,.png,.jpg}
\usepackage{thumbpdf}
\usepackage{color}
\definecolor{darkgreen}{rgb}{0,0.5,0}
\definecolor{darkblue}{rgb}{0,0,0.5}
\definecolor{brown}{rgb}{0.98,0.92,0.73}
\definecolor{red}{rgb}{1,0,0}
\definecolor{yellow}{rgb}{1,1,0}
\definecolor{blue}{rgb}{0,0,1}
\definecolor{green}{rgb}{0,1,0}
\definecolor{purple}{rgb}{1,0,1}
\definecolor{gray}{rgb}{0.8,0.8,0.8}
\definecolor{black}{rgb}{0,0,0}
\definecolor{white}{rgb}{1,1,1}
\definecolor{gold}{rgb}{1.,0.84,0.}

\graphicspath{{~/Documents/Images//}{Figures//}}

\usepackage{amsmath}
\usepackage{wasysym}

\usepackage{booktabs}

\usepackage{lineno}

\usepackage{pdfsync}

\usepackage{doi}

\usepackage{siunitx}

\usepackage{nicefrac}

\usepackage{subfigure}

\def\bs{\boldsymbol}

\def\bs{\boldsymbol}

\usepackage[paper=a4paper,total={170mm,241mm},top=20mm,left=20mm,columnsep=8mm]{geometry}\usepackage{siunitx}
\usepackage{nicefrac}
\def\smallfigurewidth{0.6\columnwidth}
\def\figurewidth{0.8\columnwidth}
\def\widefigurewidth{\columnwidth}

\newif\ifcmtr
\cmtrfalse
\ifcmtr 
\newcommand{\cmtr}[1]{ %
   [\color{red} \textbf{#1} \normalcolor]%
}%
\else
\newcommand{\cmtr}[1]{ %
}%
\fi

\newif\ifcmtrj
\cmtrjtrue%
\ifcmtrj 
\newcommand{\cmtrj}[1]{ %
   [\color{green} \textbf{#1} \normalcolor]%
}%
\else
\newcommand{\cmtrj}[1]{ %
}%
\fi

\begin{document}
\title{Temperature dependence of the energy barrier and switching field\\ of
  sub-micron magnetic islands with perpendicular anisotropy}

\date{\today}

\author{Jeroen de Vries$^1$}
\author{Thijs Bolhuis$^1$}
\author{Leon Abelmann$^{1,2}$}
\affiliation{$^1$MESA$^+$ Research Institute, University of
 Twente, The Netherlands\\$^2$KIST Europe,
 Saarbr\"ucken, Germany\\\underline{l.abelmann@kist-europe.de}}

\begin{abstract}
  Using the highly sensitive anomalous Hall effect (AHE) we have been
  able to measure the reversal of a single magnetic island, of
  diameter \SI{220}{\nano\meter}, in an array consisting of more than
  80 of those islands. By repeatedly traversing the hysteresis loop,
  we measured the thermally induced fluctuation of the switching
  field of the islands at the lower and higher ends of the switching
  field distribution. Based on a novel easy-to-use model, we
  determined the switching field in the absence of thermal activation,
  and the energy barrier in the absence of an external field from
  these fluctuations.  By measuring the reversal of individual dots in
  the array as a function of temperature, we extrapolated the
  switching field and energy barrier down to \SI{0}{K}. The
  extrapolated values are not identical to those obtained from the
  fluctation of the switching field at room temperature, because the
  properties of the magnetic material are temperature dependent. As a
  result, extrapolating from temperature dependent measurements
  overestimates the energy barrier by more than a factor of two. To
  determine fundamental parameters of the energy barrier between
  magnetisation states, measuring the fluctuation of the reversal
  field at the temperature of application is therefore to be
  preferred. This is of primary importance to applications in data
  storagea and magnetic logic. For instance in fast switching, where
  the switching field in the absence of thermal activation plays a
  major role, or in long term data stability, which is determined by
  the energy barrier in the absence of an external field.
\end{abstract}

\maketitle 

\section{Introduction}

Sufficiently small magnetic elements have only two stable
magnetisation states, separated by a an energy barrier. At finite
temperature, the system can spontaneously jump from one state to the
other. If we lower the energy barrier by an external magnetic field,
the time before jumping reduces until it is limited by spin
dynamics.

To understand magnetisation reversal, for instance for application
in non-volatile data storage, we need to know the height of the energy
barrier and how it changes with an externally applied field. We are
particularly interested in a) the height of the energy barrier in the
absence of an external field and b) the field required for reversal in
the absence of thermal energy. These fundamentally important
properties of the energy barrier are surprisingly difficult
to determine experimentally. We have two parameters to play
with: temperature and time.

In temperature dependent measurements, one measures hysteresis loops
over a wide temperature range~\cite{Schuh2001,Wernsdorfer1997}. From
the temperature dependence of the switching field, one can calculate
the height of the energy barrier. This method, however, suffers from
the fact that material properties are temperature dependent. As we
show in this paper, an estimate of the energy barrier from
extrapolation of temperature dependent measurements can lead to large
errors.

It is therefore in principle better to determine the properties of the
energy barrier at room temperature, which is usually done by observing
the hysteresis loop under different field ramp
rates~\cite{Wierenga2000,Sharrock1994,Victora1989,Sharrock1981}.  However, this
is experimentally challenging. On the low side, field ramp
rates are limited by the total time required for the measurement. On
the high side, the ramp rate is limited by the power required to build
up the field in a short time. Usually a combination of equipment is
used, where the low ramp rates (0.01 to \SI{10}{mT/s}) are measured in a
vibrating sample magnetometer (VSM)~\cite{Collocott2012} and the high ramp rates
(\si{kT/s} to \si{MT/s}) using pulsed fields~\cite{Ito2008}. The intermediate
region is difficult to address.

Non-volatile data storage materials need to retain the magnetisation
state over years, and therefore require high energy barriers. In order
to avoid excessively high temperatures or long measurement times, both
the temperature and time dependent measurements need to be combined with
an external magnetic field lowering the energy barrier. The only way
to determine the energy barrier in the absence of a magnetic field is
to employ a model relating the height of the energy barrier to the
external field. An analytical model for the field dependence of the energy barrier
of sub-micron magnetic discs with perpendicular anisotropy is
discussed in the theoretical section of this paper.

Recently, we proposed a novel method to determine the energy barrier
at room temperature~\cite{Engelen2010ahe, DeVries2013}. Rather than
increasing the field at different ramp rates, we repeatedly reverse
the magnetic islands at the same ramp rate. On every attempt, the
contribution of the thermal energy in the system will be slightly
different, leading to a fluctuation of the switching field between
attempts. With many measurements, we obtain a thermal switching field
distribution (SFD$_T$), from which the switching field in the absence
of thermal fluctuations ($H_\text{n}^\text{0}$) and the energy barrier
in the absence of an external field ($\Delta U_0$) can be
determined~\footnote{We previously used $H_\text{s}^\text{0}$ and $E_0$
  for these parameters}. A similar approach was used to study domain
wall pinning by Yun \emph{et al.}~\cite{Yun2014}.

In this paper, we extended our anomalous Hall effect (AHE) setup with
a cryostat to enable measurements in a temperature range from 10 to
\SI{300}{K}. This allows us to compare our novel statistical method with
temperature dependent measurements of the switching field. To
illustrate that indeed the temperature dependent method suffers from
the changes in material properties, we measured the temperature dependence
of the saturation magnetisation ($M_\text{s}(T)$) and effective
anisotropy ($K_\text{eff}(T)$) by VSM and torque magnetometery.

Our modified AHE setup allows us to perform repeated experiments at
\SI{10}{K} as well as at room temperature. In this way we can determine
the changes in the energy barrier with temperature, which can be related
to changes in the nucleation volume and wall energies using our novel
analytical model.

These observations are of importance for applications using patterned
magnetic elements. One example is bit patterned magnetic media, which
is one of the possible solutions to postpone the superparamagnetic limit
that current hard disk technology is approaching. The height of the
energy barrier, and its relation to the external magnetic field,
determines the long term stability of the data. A problem that still
needs to be overcome is the large variation in the required switching
field between elements~\cite{Chen2010}. This switching field distribution is probably
caused by an intrinsic anisotropy distribution that is already
present before patterning \cite{Thomson2006,Shaw2007}
  Our method
provides insight into the variation of the energy barriers between the islands,
and therefore indirectly into the variation in the anisotropy.

A second example is the patterned magnetic elements in magnetic random
access memories (MRAM) or magnetic logic, which suffer from the thermally activated
variations in the switching field~\cite{Breitkreutz2015,Wang2004}. Our method allows
the determination of the switching field in the absense of thermal
fluctuation at room temperature.  To study ultra-fast switching, this
value needs to be known in order to determine the increase in switching field
due to reversal dynamics~\cite{Wang2008a,Banholzer2011}.

\begin{figure}
    \centering
    \includegraphics[width=\figurewidth]{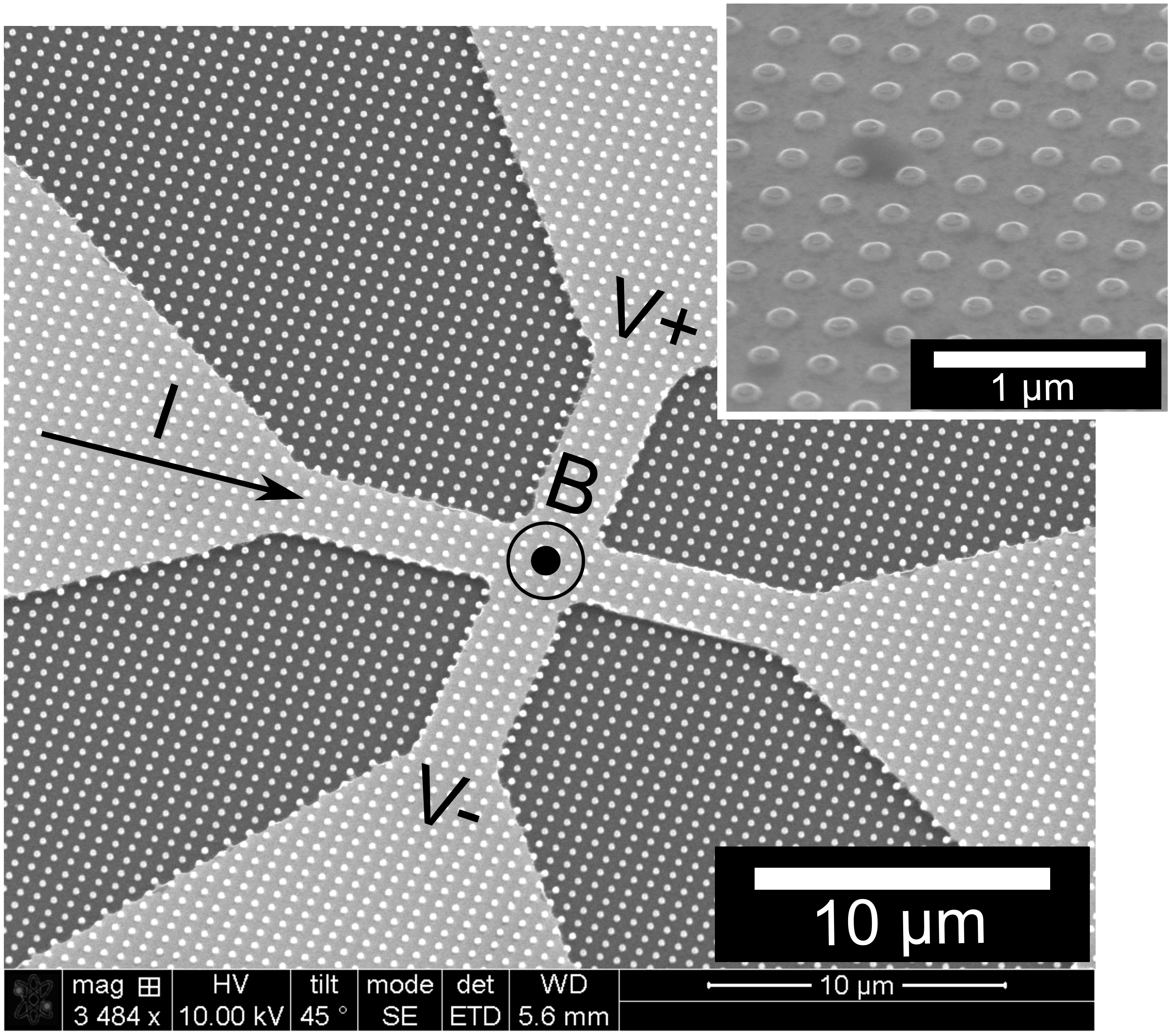}
    \caption{SEM picture of a Hall cross structure with magnetic
     islands on top, indicating the direction of the current ($I$),
     magnetic field ($B$) and measured Hall voltage ($V$). The inset
     shows a zoom of the area with magnetic islands before patterning
     of the Hall-cross.}
    \label{fig:HallCross}
\end{figure}

\section{Theory}

\subsection{Switching field and energy barrier}
From statistical measurements of the switching field of a single
island it is possible to determine the energy barrier in the absence
of an external field ($\Delta U$) and the switching field in the absence
of thermal fluctuations ($H_\text{n}^0$). In the following, we
derive the basic theory for linking these values to the measured distribution
of the switching field of an individual island.

\subsubsection{Thermally induced reversal}
Consider a system, like a single domain magnetic island, that has two
energy minima, separated by an energy barrier of finite height
$\Delta U$. Due to thermal fluctuations, there is a chance that the
system jumps between the energy minima. We assume this probability can
be described by Arrhenius statistics.  At time $\tau$=\SI{0}{s}, the
system is in one energy minimum. The probability that the system has jumped
to the other energy minimum increases with time:

\begin{align}
\label{eq:4}
P_\text{sw}\left(\tau\right)&=1-\exp\left(-\tau/\tau_0\right),\\
\tau_0&=\frac{1}{f_0}\exp\left(\frac{\Delta U(H,T)}{kT}\right)
\end{align}

where $f_0$ is the frequency [Hz] at which the system tries to attempt
to overcome the energy barrier, $k$ is Boltzmann's constant (\SI{1.38e-23}{J/K}) and $T$ the
temperature [K].

When taking a hysteresis loop of our magnetic islands, we slowly ramp
up the field from some negative field value, where all islands are in
the same state, $-H_\text{sat}$ in small steps $\Delta H$ and monitor
the reversal of the magnetisation in the islands after each step for a
waiting time $\Delta \tau$. We assume that the waiting time is short
enough to make it very unlikely that there will be multiple reversals,
back and forth between the energy minima. In this case, the
probability that the magnetisation in the island switches at a field value $H$ is the
chance that it switches within the waiting time (Equation~\ref{eq:4}),
multiplied by the chance that it has not yet switched before,

\begin{multline}
  \mathcal{P}_\text{sw}\left( H, \Delta \tau \right)=\\
  P_\text{sw}(\Delta \tau) \left( 1-
    \int_{-H_\text{sat}}^H  p_\text{sw}\left(H',\Delta \tau\right)dH'\right).
\end{multline}

In the above, $p_\text{sw}$ is the corresponding probability density
function [m/A]. This implicit equation can be reformulated explicitly
if we assume the field steps are so small that we can define a
continuous field ramp rate $R=\Delta H / \Delta \tau$ [\SI{}{\ampere\per\meter\per\second}]. In that
case, the probably density function becomes~\cite{Engelen2010ahe}

\begin{multline}
\label{eq:7}
 p_\text{sw}\left(H,T\right)=\frac{f_0}{R}\text{exp}\left(\frac{-\Delta
     U\left(H,T\right)}{kT}\right)\\
 \times  \exp \left[-\frac{f_0}{R} \int_{-H_\text{sat}}^H
   \exp\left(\frac{-\Delta U(H',T)}{kT}\right) dH'\right].
\end{multline}

We explicitly take into account that the energy barrier is dependent
on the temperature at which the distribution is measured, because of
the variation with temperature of the magnetic properties of the
material. However, the crucial information required is the exact way
in which the energy barrier, $\Delta U$, decreases with a decrease in
strength of the applied field. The relation between the energy barrier
and the applied field depends strongly on the way the islands reverse
their magnetisation direction. In the following we will describe two
extreme models: coherent rotation and domain wall creation and
propagation.

\subsubsection{Field dependent energy barrier: Coherent rotation}
\begin{figure}
    \centering
    \includegraphics[width=\smallfigurewidth]{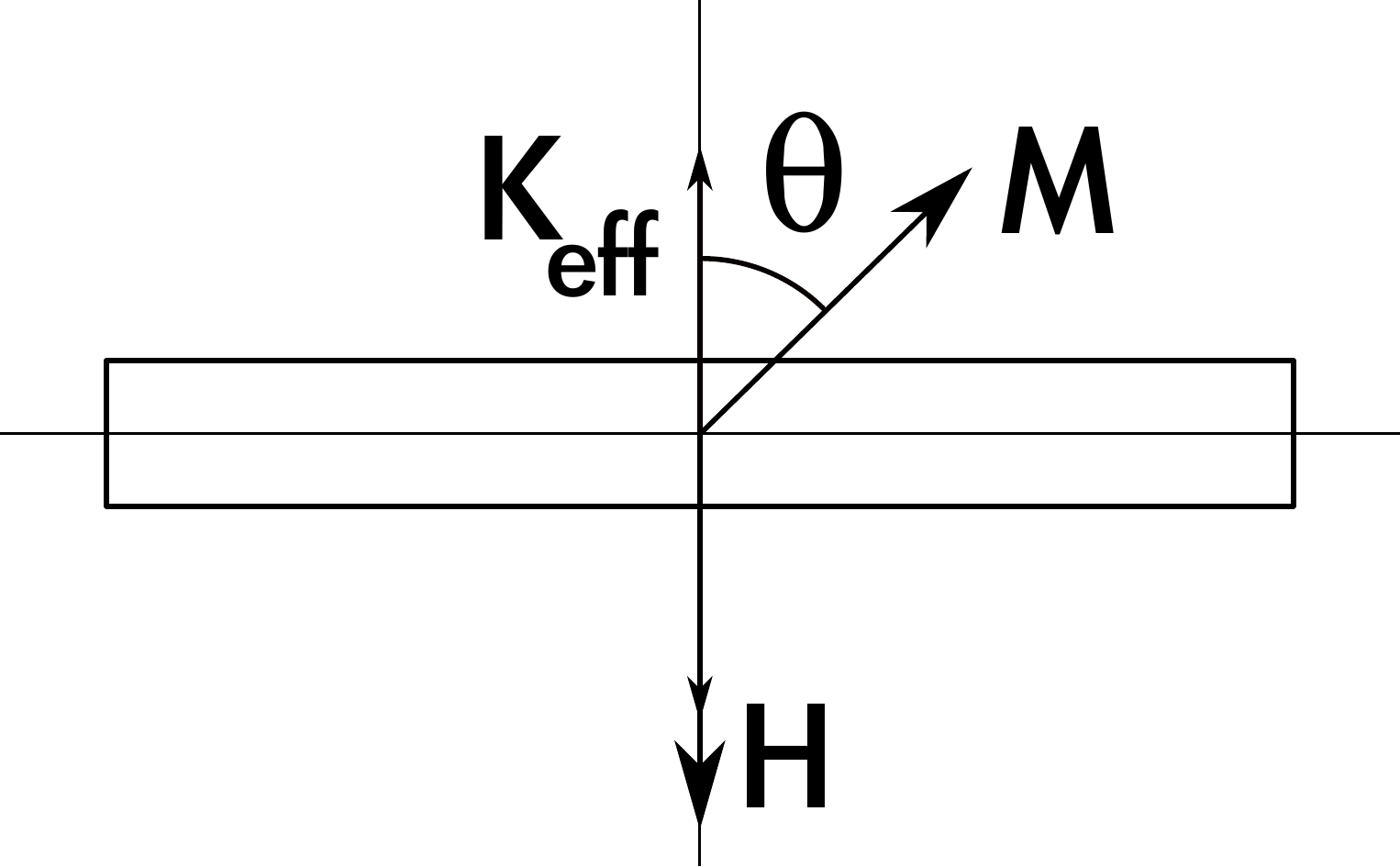}
    \caption{The simplest model discussed assumes that the
      magnetisation $\bs{M}$ coherently rotates away from the easy
      axis $\bs{K}_\text{eff}$ over an angle $\theta$ if the applied field $\bs{H}$ is
      increased.}
    \label{fig:coherent}
\end{figure}

In the coherent rotation model (Stoner--Wohlfarth), we assume that the
spins in the island remain parallel during rotation. This model is
well described~\cite{Stoner1991}, but repeated here since it defines
an upper limit to the switching field that should be compared to
alternative models. The model assumes an effective anisotropy
$K_\text{eff}$ [J], which tries to align the spins parallel to the easy
axis at an angle $\theta$, and an external field $H$ that tries to
align the spins along the field direction
(Figure~\ref{fig:coherent}). The total energy of the system is the sum
of the anisotropy and the external field energy

\begin{equation}
U_\text{I}=\left(K_\text{eff} \sin^2\theta +
  \mu_0M_\text{s}H\cos\theta\right)V\text{ [J]}
\end{equation}

Where $V$ is the sample volume, $M_\text{s}$ the sample's saturation
magnetisation [A/m], and $\mu_0$ the vacuum permeability
[$4\cdot10^{-7}\pi$ Tm/A]. The extrema in the energy function
can be found by equating to zero the derivative of the energy with respect to
$\theta$, which leads to $\theta=0$ for the minimum and 

\begin{equation}
  \label{eq:3}
  \cos(\theta_\text{max})=\frac{\mu_0M_\text{s}H}{2K_\text{eff}}
\end{equation}

for the maximum energy. The energy barrier is the difference between
the maximum and minimum,

\begin{equation}
  \label{eq:5}
  \Delta
  U_\text{I}(H)=K_\text{eff}V\left(1-\frac{H}{H^0_\text{I,n}}\right)^2 
\end{equation}

with the switching field

\begin{equation}
   H^0_\text{I,n}=\frac{2K_\text{eff}}{\mu_0M_\text{s}}\text{ [A/m]}.
\end{equation}

We use the upper index \num{0} to indicate the switching field in the
absence of thermal fluctuation.

Since all spins in the island switch in unison, the switching volume
$V_\text{sw}$ is equal to the island volume~$V$.

\subsubsection{Field dependent energy barrier: Domain wall motion}
\label{sec:field-depend-energy}
For the \SI{220}{nm} islands that we measured, the coherent rotation model is
too coarse an aproximation.
It is more likely that reversal starts in a small region with low
anisotropy, followed by the propagation of a domain wall through the
island~\cite{Hu2005,Lau2011,Delalande2012}. The theoretical background
for this reversal mechanism has been beautifully explained by Adam and
co-workers~\cite{Adam2012} in their bubble growth model. Their
approach, however, lacks the simplicity of the Stoner--Wohlfarth model. We
therefore modified their circular geometry to a square shape, while
keeping the essence of their model. In contrast to the approach by
Adam, this simplified model leads to a closed form solution. Even though our
islands are circular, not square, the predicted trends will be very similar. In
the following, we describe this \emph{diamond} model, and discuss its
implications for a wall energy density that is either constant or
varies with position.

\paragraph{The diamond model}

\begin{figure}
    \centering
    \includegraphics[width=0.35\columnwidth]{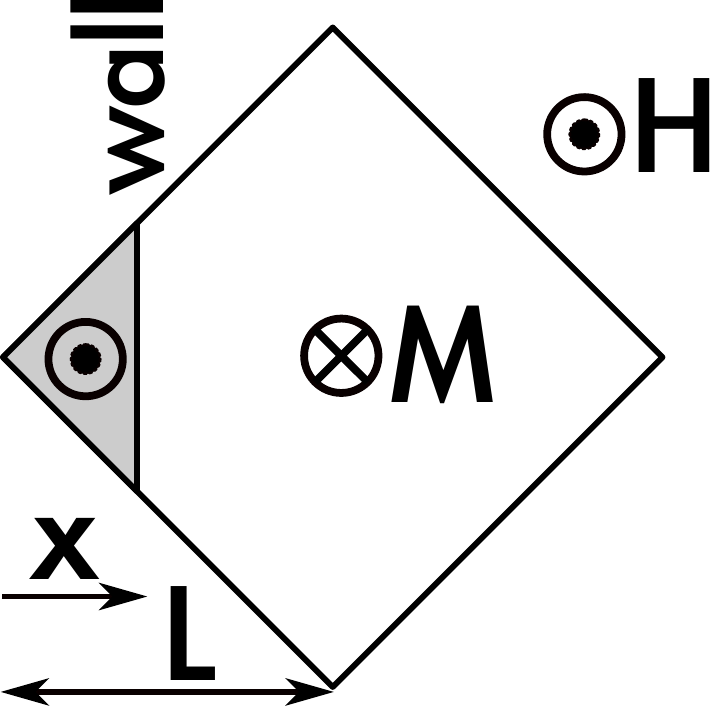}
    \caption{Rather than assuming coherent rotation, it is more
      realistic to assume creation and propagation of a domain
      wall. We assume a square island of area $2L^2$, into which a domain wall propagates 
      from a corner over a distance $x$.}
    \label{fig:diamond}
\end{figure}

Consider a square magnetic element of thickness $t$ and area $2L^2$,
with an out of plane easy axis (Figure~\ref{fig:diamond}). The
magnetisation in the element is pointing downwards, and an opposing
field $H$ is applied. Reversal starts by introduction of a domain wall
at position $x=0$. The total energy of the system is the sum of the
wall energy, proportional to the wall length and the wall energy
density $\sigma$ [J/m$^2$], with the external field energy, which is
proportional to the area of the reversed domain and the external
magnetic field. For $x\le L$,

\begin{align}
  \label{eq:6}
  U_\text{II}&= \nonumber\\
  &2x\sigma t-2\mu_0M_\text{s}Ht\left(x^2-L^2\right) 
  &\text{for } x\le L \nonumber \\
  &2(2L-x)\sigma t-2\mu_0M_\text{s}Ht\left(x^2-L^2\right) 
  &\text{for } x> L
\end{align}

The force on the domain wall is the negative derivative of the energy
with respect to the wall position $x$. Nucleation of a domain occurs when the
force changes sign, which is when the derivative passes zero. The
field at which this occurs, $H^0_\text{II,n}$, is defined as the
nucleation field. In the absence of pinning sites, so in a perfectly
homogeneous material, the domain wall will continue to propagate until
the magnetisation in the island is reversed completely. In this case
the nucleation field is equal to the switching field. In reality the
domain wall might be trapped~\cite{Delalande2012}, and next to a
nucleation field there  will be one or several domain wall depinning
fields before the island switches. This case is not considered
here.

\paragraph{Constant wall energy}

We first consider the wall energy density to be independent of 
position ($\sigma=\sigma_0$).  Equating to zero the derivative of the energy
with respect to $x$, we obtain, for the wall position at which
nuclation occurs,

\begin{equation}
  \label{eq:8}
  x_\text{max}=\frac{\sigma_0}{2\mu_0M_\text{s}H}
\end{equation}

We assume $x_\text{max} \leq L$,  which implies that Equation~\ref{eq:8}
is only valid for

\begin{equation}
  \label{eq:9}
  H \ge \frac{\sigma_0}{2 \mu_0 M_\text{s}L}= H_\text{L}
\end{equation}

For $H<H_\text{L}$, nucleation will occur if the wall reaches the
widest part of the diamond, so $x_\text{max}=L$. 

\begin{figure}
    \centering
    \includegraphics[width=\widefigurewidth]{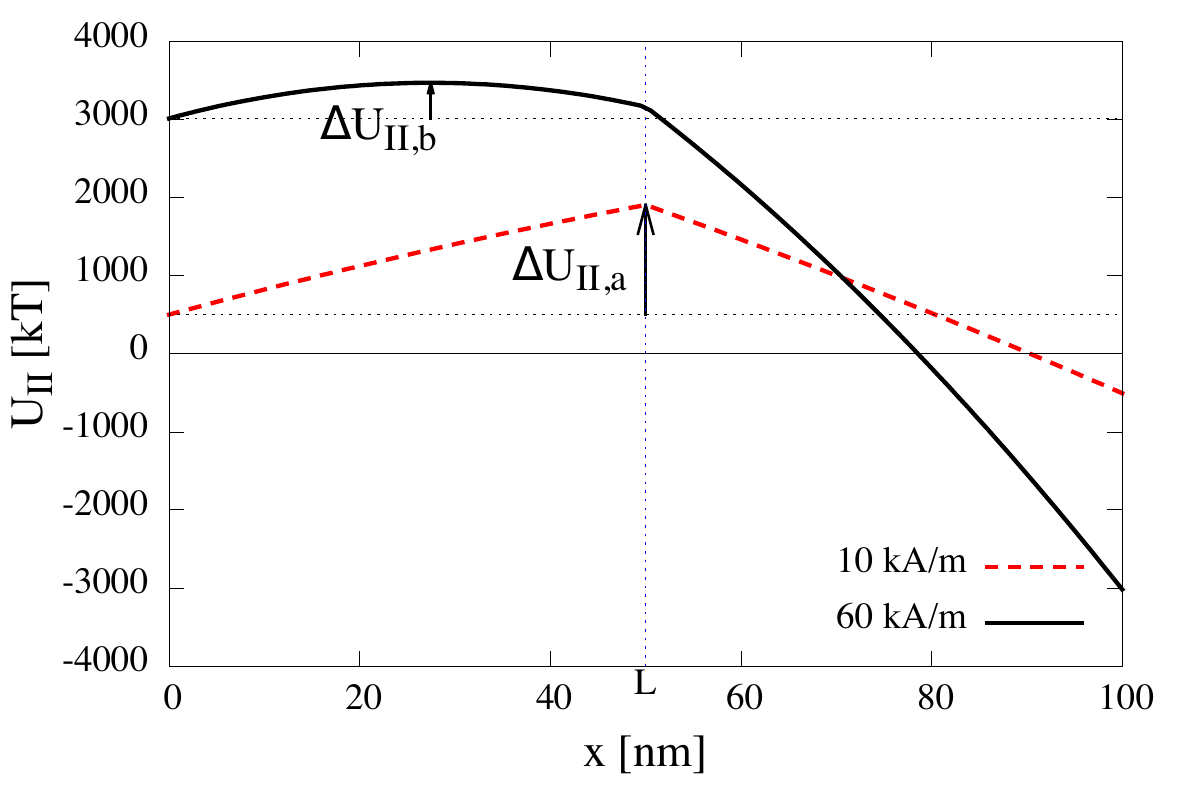}
    \caption{Energy (in units of $kT$ at \SI{300}{K}) versus wall position, for a field below
      $H_\text{L}$ where nucleation occurs at $x=L$ (label ``a''), and
      for a field above $H_\text{L}$ where nucleation occurs at
      $x<L$ (label ``b''). To generate this plot, we have used the values given in
      Table~\ref{tab:exampleparameters}.}
    \label{fig:U2vsX}
\end{figure}

Figure~\ref{fig:U2vsX} shows the energy versus the wall position, for
both situations. At low fields (\SI{10}{kA/m} in the graph), nucleation
occurs when the wall reaches the widest part of the triangle, i.e., at
$x_\text{max}=L$. At high fields $x_\text{max}<L$. As can be seen, the
height of the energy barrier, $\Delta U$, depends on the location of the maximum
energy, $x_\text{max}$. We must consider two cases.

\begin{table}
  \begin{ruledtabular}
  \caption{Parameters used to generate the graphs of
    Figures~\ref{fig:U2vsX} to~\ref{fig:PallvsH}. }
  \label{tab:exampleparameters}
  \begin{tabular}[t]{llll}
    $M_\text{s}$& \SI{829}{kA/m} &     $kT$ & \SI{25.84}{meV}\\
    $\sigma_0$ & \SI{3.43}{mJ/m^2} & $f_0$ & \SI{e9}{Hz}\\
    $ K_\text{eff}$ & \SI{386}{kJ/m^3} & $R$ & \SI{50}{A/ms}\\
    $L$ & \SI{50}{nm} & $H_\text{L}$ & \SI{32.9}{kA/m}\\
    $t$  & \SI{20}{nm} & $H^0_\text{I,n}$ & \SI{741}{kA/m}\\
    $w$ & \SI{16}{nm} & $H^0_\text{III,n}$ & \SI{206}{kA/m}\\
  \end{tabular}
  \end{ruledtabular}
\end{table}

\subparagraph{Regime A, $x_\text{max}=L$, $H \le H_\text{L}$}
At low fields, $H \le H_\text{L}$, the wall must propagate all the way to the widest part of the
diamond for nucleation. In this case, $x_\text{max}=L$ and the height of the energy barrier is

\begin{align}
  \label{eq:10}
  \Delta U_\text{IIa}(H)&=U_\text{II}(L)-U_\text{II}(0)\nonumber\\
                        &=2L\sigma_0t\left(1-\frac{H}{H^0_\text{II,n}}\right).
\end{align}

\begin{figure}
    \centering
    \includegraphics[width=\widefigurewidth]{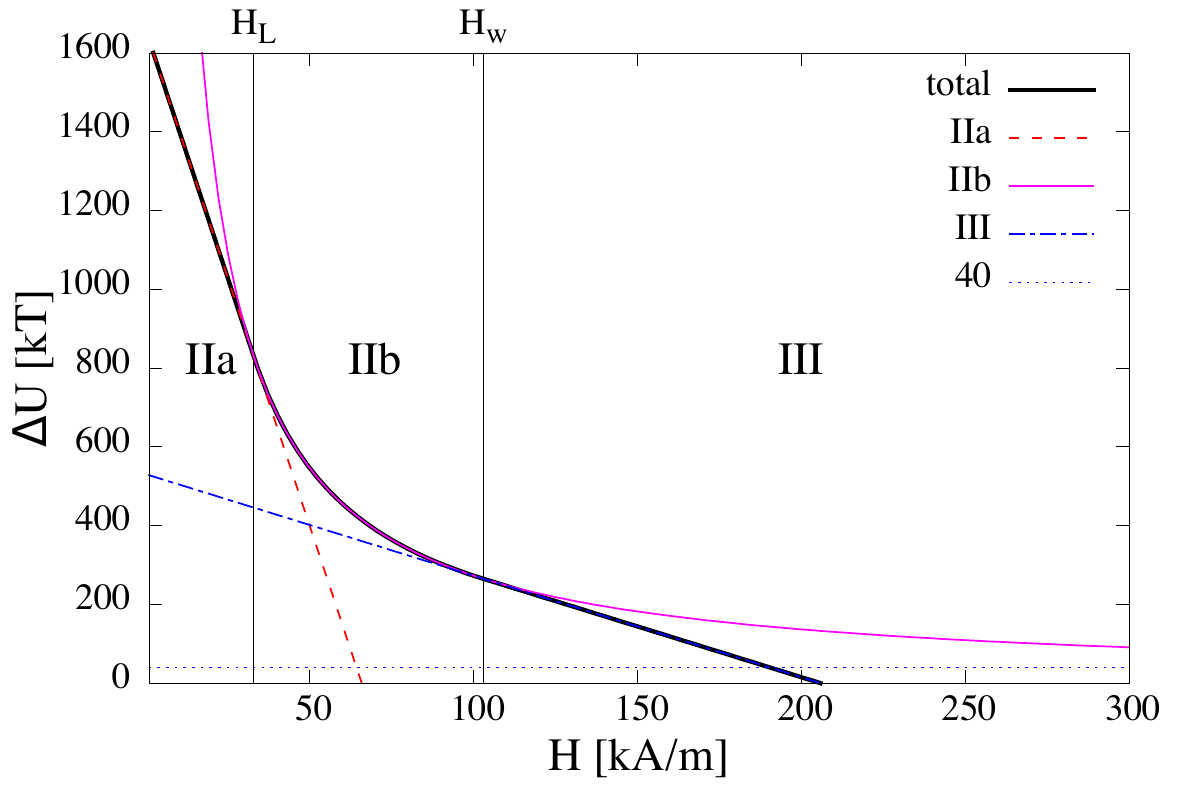}
    \caption{Energy barrier (in units of $kT$ at \SI{300}{K}) versus
      applied field. For fields below $H_\text{L}$, nucleation occurs
      when the wall reaches the widest part of the diamond (regime
      ``IIa''). For larger fields, nucleation is reached before the
      wall reaches the widest part (regime
      ``IIb''). But in model IIb, the energy barrier never
      decreases to zero (magenta line). Only if the domain wall energy
      density is assumed to increase linearly from zero from the edge
      of the island over a distance $w$, is a reasonable switching field
      obtained (model III, which is valid for $H>H_\text{w}$). The
      dotted line at \SI{40}{kT} indicates the energy barrier which can
      generally not be overcome in normal experimental conditions. To
      generate this plot, we have taken the values given in
      Table~\ref{tab:exampleparameters}.}
    \label{fig:DU2vsH}
\end{figure}

The energy barrier is plotted as a function of the field in
Figure~\ref{fig:DU2vsH}, we are considering region ``IIa'' here. The
red dashed line shows the extrapolation of the energy barrier for
values above $H_\text{L}$.

The nucleation field in the absence of thermal activation is equal to

\begin{equation}
  \label{eq:11}
  H^0_\text{IIa,n}=\frac{\sigma_0}{\mu_0M_\text{s}L},
\end{equation}

which is the intersection of the red dashed line with the $H$-axis. This nucleation
field is twice the value of $H_\text{L}$, so if there is no thermal
activation, this nucleation mechanism will never occur. The maximum
field at which reversal in this regime can take place is at $H=H_\text{L}$, where
the energy barrier $\Delta U_\text{IIa}=L\sigma_0t$. Using realistic
values (Table~\ref{tab:exampleparameters}), this energy still is
around \SI{800}{kT}, so for our situation nucleation will not
occur in regime IIa. The next question is therefore whether nucleation
can occur at all before the wall reaches the widest part of the diamond.

\subparagraph{Regime B, $x_\text{max}<L$, $H>H_\text{L}$}
At applied fields above $H_\text{L}$, nucleation occurs before the wall reaches the widest part of
the diamond, $x<L$, and the energy barier equals

\begin{equation}
  \label{eq:12}
  \Delta U_\text{IIb}(H)=U_\text{II}(x_\text{max})-U_\text{II}(0)=
  \frac{\sigma_0^2t}{2\mu_oM_\text{s}H}.
\end{equation}

The decrease in the energy barrier with increasing applied field
strength is shown in Figure~\ref{fig:DU2vsH}, indicated by the solid
magenta line ``IIb''. The energy barrier never decreases to zero, so
$H^0_\text{IIb,n}=\infty$. By thermal activation however, nucleation
can occur at a finite field, in which case the switching volume is

\begin{equation}
  \label{eq:12b}
  V_\text{sw}=x_\text{max}^2t=\frac{\sigma_0^2t}{4\mu_0M_\text{s}^2H^2}.
\end{equation}

\begin{figure}
    \centering
    \includegraphics[width=\widefigurewidth]{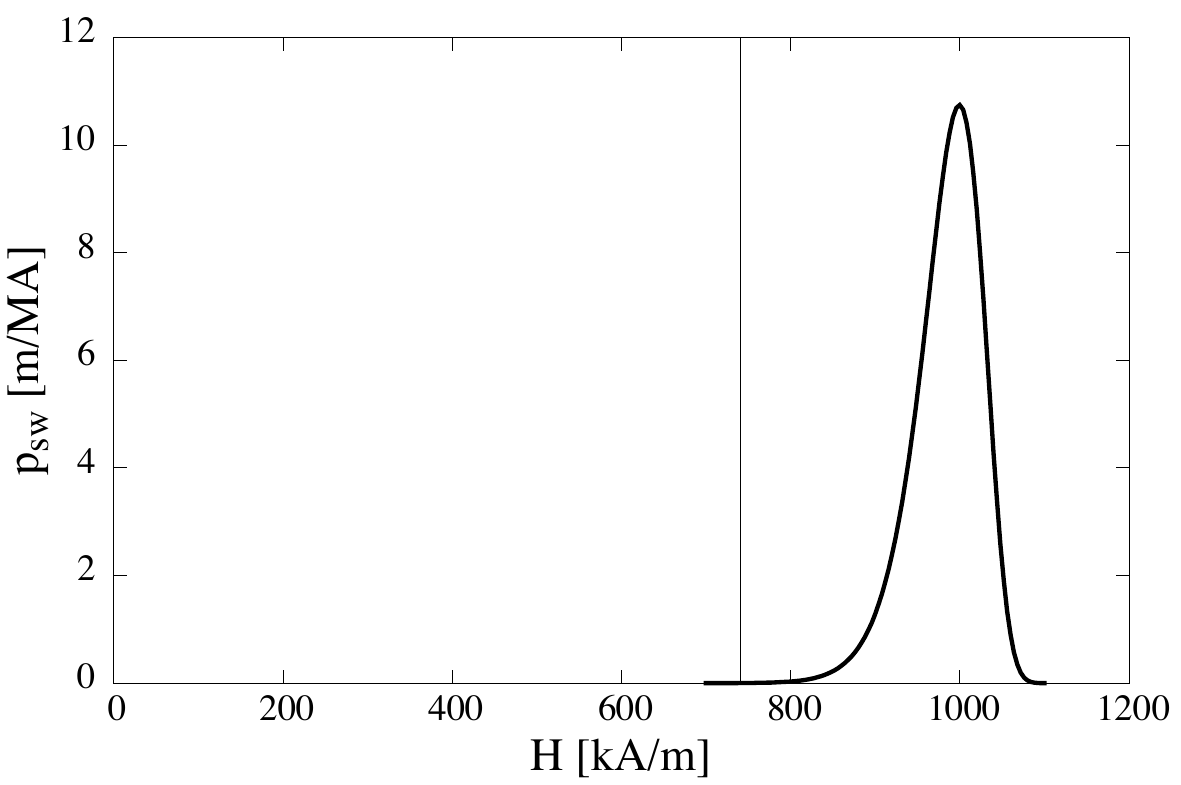}
    \caption{Thermally activated switching field distribution for a
      domain wall movement model using a constant wall energy. The
      switching fields for this oversimplified model are
      unrealistically high, since they are higher than the nucleation
      field for coherent rotation in the absence of thermal activation
      (line at \SI{740}{kA/m}).}
    \label{fig:P2vsH}
\end{figure}

Figure~\ref{fig:P2vsH} shows the calculated switching field
distribution for model II at room temperature. All switching fields
are above \SI{800}{kA/m}. The coherent rotation model would give a
switching field of \SI{741}{kA/m}, which would therefore be the
preferred reversal mode, similar to the reversal model discussed in~\cite{Adam2012}. For
our set of parameters, neither regime ``IIa'' or ``IIb'' results in
realistic switching fields and this model must be discarded.

\paragraph{Linearly increasing wall energy}

\begin{figure}
    \centering
    \includegraphics[width=0.5\columnwidth]{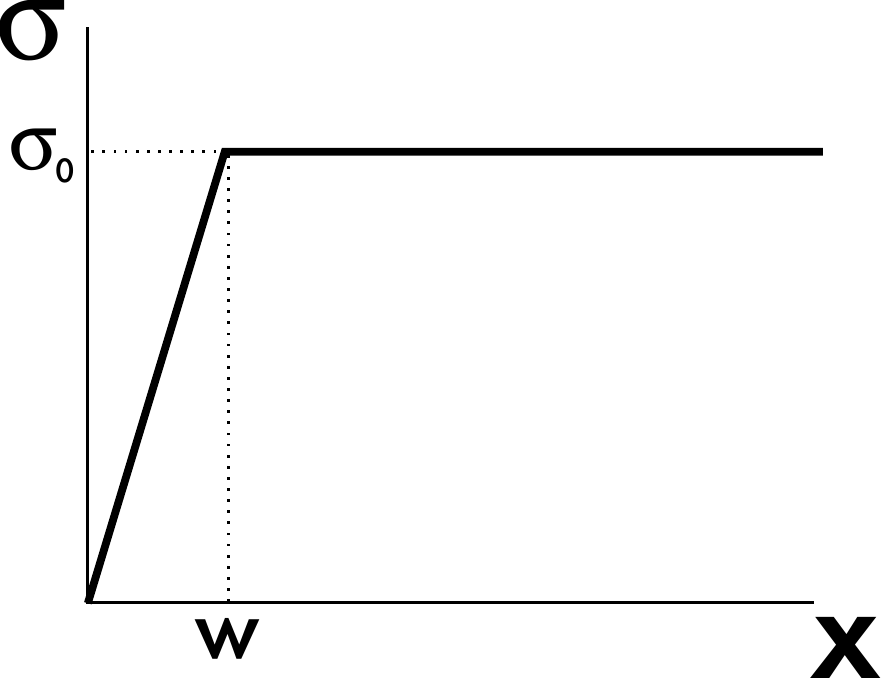}
    \caption{In the final version of the diamond model, we assume that the
      domain wall energy increases linearly with distance, as it
      propagates into the square island, up to $x=w$ (see Figure~\ref{fig:diamond}).}
    \label{fig:linearsigma}
\end{figure}

Since the constant wall energy model above leads to unrealistic values
for the nucleation field, we assume a wall energy that increases with
position. The simplest assumption is a linear increase, over a
distance $w$ (Figure~\ref{fig:linearsigma}).

When $x<w$, the domain wall is in the linearly increasing part, and the
energy as a function of wall position is

\begin{align}
  \label{eq:13}
  U_\text{III}&=2tx^2\frac{\sigma_0}{w}-2\mu_0M_\text{s}tH(x^2-L^2)\nonumber\\
                  &=2t\left(\frac{\sigma_0}{w}-\mu_0M_\text{s}H\right) x^2 +\text{constant}
\end{align}

For $x>w$, the energy is as before in model II. By equating to zero the
derivative with respect to $x$, we obtain the nucleation field
in the absence of thermal activation:

\begin{equation}
  \label{eq:14}
  H^0_\text{III,n}=\frac{\sigma_0}{\mu_0M_\text{s}w}.
\end{equation}

The result is similar to model IIa (Equation~\ref{eq:10}
and~\ref{eq:11}), with $L$ replaced by $w$. Following the
analogue to the line of argument for model IIa, we can discriminate three
regions, separated by $H_\text{L}$ (Equation~\eqref{eq:9}) and 

\begin{equation}
  \label{eq:16}
  H_\text{w}=\frac{\sigma_0}{2\mu_0M_\text{s}w}=\frac{H^0_\text{III,n}}{2}.
\end{equation}

\begin{figure}
    \centering
    \includegraphics[width=\widefigurewidth]{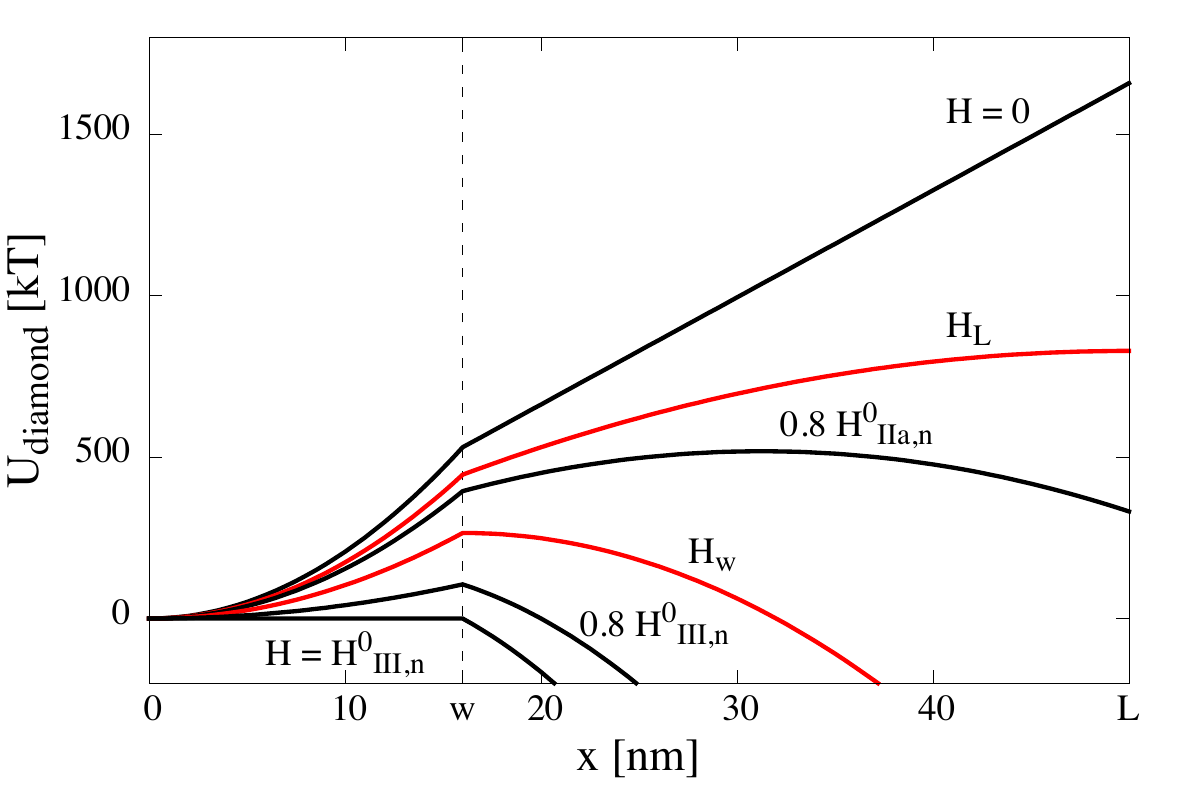}
    \caption{Energy as a function of wall position, assuming a linear
      increase of wall energy density for $x<w$. For $H<H_\text{L}$,
      the maximum energy is found when the wall reaches the widest
      part of the diamond ($x=L$). For $H_\text{L}<H<H_\text{w}$, the
      maximum energy lies between $x=w$ and $L$. For $H>H_\text{w}$,
      the energy barrier is located at $x=w$, until it disappears at
      the nucleation field $H^0_\text{III,n}$.}
    \label{fig:U3vsX}
\end{figure}

An example of the energy function is shown in Figure~\ref{fig:U3vsX}. For $H<H_\text{w}$, the maximum energy is found at $w<x<L$, and we have
the situation of model II. As discussed before, nucleation in this
regime does not occur for realistic temperatures. However, for $H_\text{w}<H<H^0_\text{III,n}$, unlike regime
IIb, the maximum energy
is always found at $x=w$. This is due to the quadratic nature of the
energy function for $x<w$. At $H=H^0_\text{III,n}$, the energy
function becomes flat, and the energy barrier disappears.

The energy barrier that is of interest to us is found at
$x_\text{max}=w$,

\begin{align}
  \label{eq:15}
  \Delta U_\text{III}(H)
  &=2t\sigma_0w\left(1-\frac{H}{H^0_\text{III,n}}\right)\\
  &=\Delta U_0 \left(1-\frac{H}{H^0_\text{III,n}}\right)
\end{align}

and is displayed together with model II in Figure~\ref{fig:DU2vsH}. In
contrast to model II, the energy barrier now decreases to zero and
nucleation can occur at realistic conditions.

Since the energy barrier is always located at $x_\text{max}=w$, the
switching volume is simply

\begin{equation}
  \label{eq:17}
  V_\text{sw}=w^2t
\end{equation}

This \emph{diamond model} has simple equations for the energy
(Equation~\ref{eq:13}) and energy barrier (Equation~\ref{eq:15}), nucleation field
(Equation~\ref{eq:14}) and volume (Equation~\ref{eq:17}). For
simplicity we will use in the remainder of this paper

\begin{equation}
H^0_\text{n}=H^0_\text{III,n}
\end{equation}

\begin{figure}
    \centering
    \includegraphics[width=\widefigurewidth]{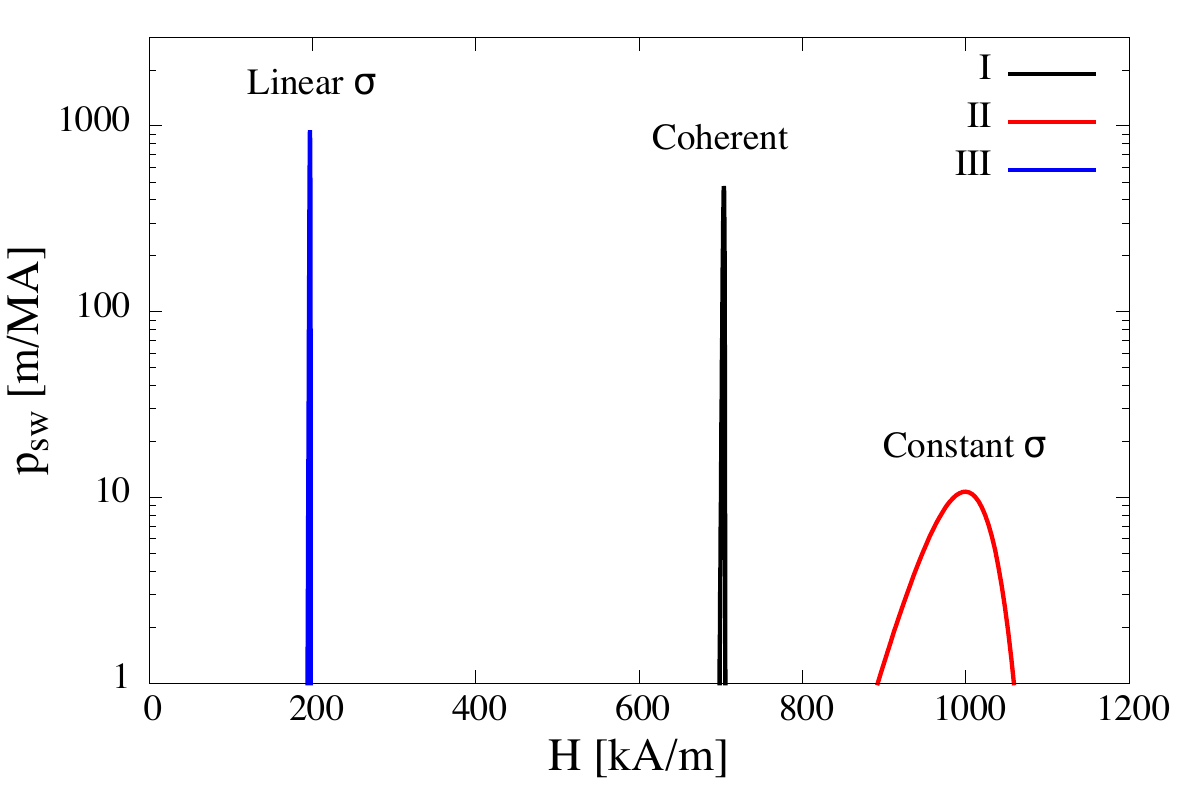}
    \includegraphics[width=\widefigurewidth]{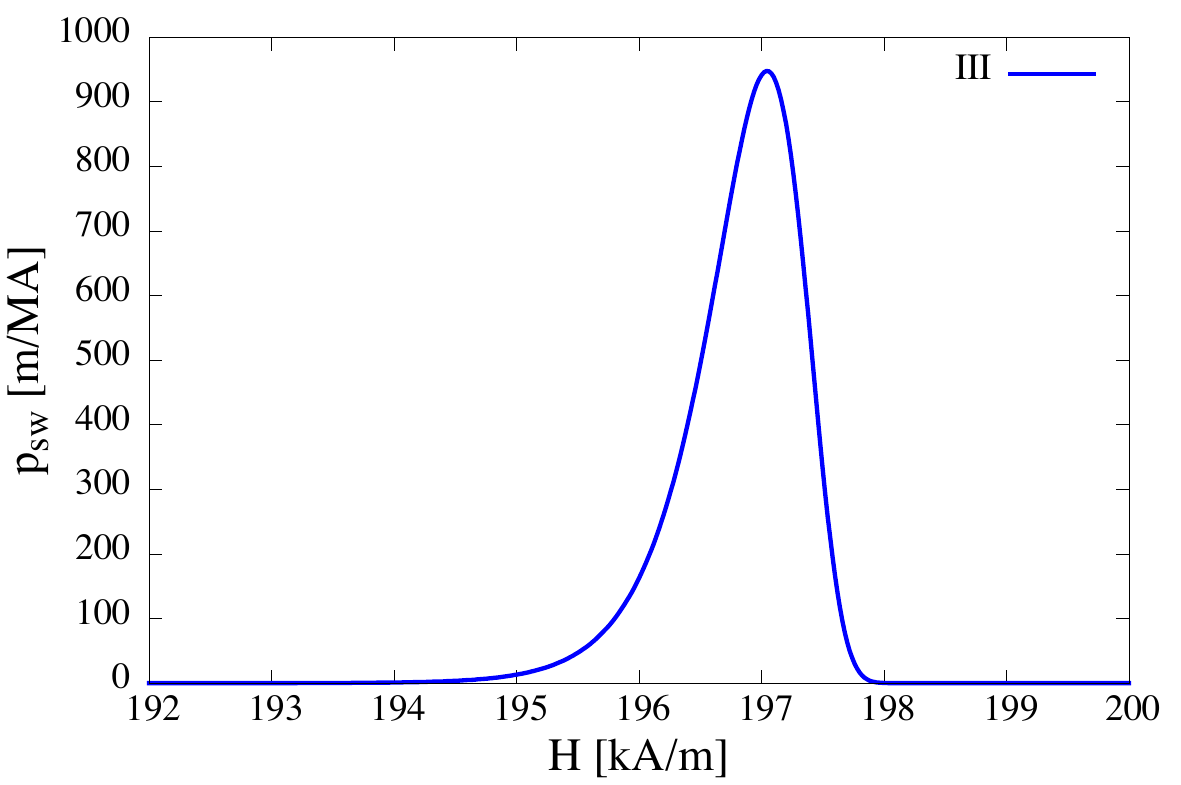}
    \caption{Top: Comparison of the thermally activated switching
      field distributions of the three different models. The model
      based on creation and subsequent domain wall movement with a
      constant wall energy $\sigma$ (red curve II) results in
      switching fields that are even higher than for a coherent
      rotation model (black curve I). However, when assuming a domain
      wall energy density that increases linearly as the wall enters the
      island, more realistic switching fields are obtained (blue curve
      III). Bottom: Zoom of the distribution for the final diamond
      model (III). See Table~\ref{tab:exampleparameters} for
      parameters used.}
    \label{fig:PallvsH}
\end{figure}

The diamond model introduces a new
parameter $w$, which is the length over which the domain wall energy
increases as the wall enters the island.  The rate of increase
in domain wall energy $\nicefrac{\sigma_0}{w}$ determines the
nucleation field. A reduction in the domain wall energy near the edge
of the island is not unrealistic. It could, for instance, be caused by a
region of reduced anisotropy at the edge of the island, due to etch
damage for instance, or by a finite wall width.  If we take a
reasonable value for $w$, \SI{16}{nm}, we obtain a quite acceptable
value for the nucleation field, as can be seen in
Figure~\ref{fig:PallvsH}, which also illustrates how, by moving
from the naive model with constant domain wall energy to the edge of
the island (red curve II) to a more realistic model with reduced
domain wall energy (blue curve III), the nucleation field can be
brought below the values for the coherent (Stoner--Wohlfarth) rotation
model (black curve I).

Under realistic experimental conditions  at room temperature, it is very unlikely that
energy barriers of more than \SI{40}{kT} are overcome by thermal
activation. As illustrated by
Figure~\ref{fig:DU2vsH}, we can safely assume that the energy barrier
decreases linearly with the applied field ($n$=1 in our earlier
work~\cite{Engelen2010ahe}). This is confirmed by
Figure~\ref{fig:PallvsH}, which shows that switching below
\SI{194}{kA/m} in this example is very unlikely. From
Figure~\ref{fig:DU2vsH} we can observe that non-linear effects start
at energy barriers above approximately \SI{400}{kT}. One would
therefore have to raise the temperature by a factor of ten before any
non-linear field dependence could be observed.

It should be noted that from the energy barrier at $H=0$, $\Delta U_0$
we can obtain the \emph{product} $\sigma_0w$ (Equation~\eqref{eq:15}),
whereas from the nucleation field in the absence of thermal
fluctuation, $H^0_\text{n}$, we can obtain the \emph{ratio}
$\nicefrac{\sigma_0}{w}$ (Equation~\eqref{eq:14}). Since both
parameters are obtained from the fit of the model to the thermal
switching field distribution curves, the domain wall energy and nucleation
volume can be determined independently.

\begin{eqnarray}
  \label{eq:18}
  w=\sqrt{\frac{\Delta U_0}{2t\mu_oM_\text{s}H_n^0}}\\
  \label{eq:18b}
  \sigma_0=\sqrt{\frac{\Delta U_0\mu_oM_\text{s}H_n^0}{2t}}
\end{eqnarray}

The nucleation volume can therefore also be written
as

\begin{equation}
  \label{eq:19}
  V_\text{sw}=w^2t=\frac{\Delta U_0}{2\mu_0M_\text{s}H^0_n}.
\end{equation}



\subsection{Temperature dependence of the magnetisation and anisotropy}

Material parameters such as saturation magnetisation and magnetic
anisotropy constant are temperature dependent. To obtain an estimate
for the magnitude of this effect when we cool down to low temperatures
in our experiments, we assume a simple Brillouin theory (taken
from~\cite{Coey2010} with $J=2$) for the temperature dependence of the
saturation magnetisation.

\begin{align}
M_\text{s}(T)/M_\text{s}(0)=\left(\frac{5}{4}\text{coth}\frac{5}{4}\chi-\frac{1}{4}\text{coth}\frac{\chi}{4}\right)
\label{eq:1}
\end{align}

where the value of $\chi$ can related to the Curie temperature $T_\text{c}$ using

\begin{align}
M_\text{s}(T)/M_\text{s}(0)=\left(\frac{T}{2T_\text{c}}\right)\chi
\label{eq:2}
\end{align}


The value of $\chi$ can be obtained graphically or by symbolic
mathematical manipulation software. We use this model to extrapolate
the measured values of $M_\text{s}(T)$ to $M_\text{s}(0)$.

We assume that the total magnetic anisotropy $K_\text{eff}$ has two
contributions: the demagnetisation energy $K_\text{d}(T)$, which is
equal to $\nicefrac{1}{2}\mu_0 M_\text{s}^2$, and an intrinsic
anisotropy $K_\text{u}(T)$. Depending on the mechanism causing the
intrinsic anisotropy, the magnetisation dependence can be on the order of
$M_\text{s}^2$ (e.g. for crystalline anisotropy~\cite{Staunton2006})
all the way up to order $M_\text{s}^3$ for pure surface
anisotropy~\cite{Asselin2010}. Therefore we model the temperature
dependence of the effective anisotropy as


\begin{equation}
K_{\text{eff}}(T)=K_\text{u}(0)\alpha^n-K_\text{d}(0)\alpha^2
\label{eq:Anis}
\end{equation}

With $n=2$ or $3$ and 

\begin{equation} \alpha =\left(\frac{5}{4}\text{coth}\frac{5}{4}\chi
-\frac{1}{4}\text{coth}\frac{\chi}{4}\right)
\label{eq:Alpha}
\end{equation}

\section{Experimental}

\subsection{Preparation of the thin magnetic film} 
The magnetic multilayer samples are prepared by cleaning
$\left<100\right>$ p-type wafers and stripping them the native oxide. A thermal
oxide layer of \SI{50}{\nano\meter} is grown by an LPCVD process. The SiO$_{2}$
acts as an insulating layer between the conducting metal layer and the
bulk silicon. A multi-target DC sputtering system is used to deposit all metal
layers in one single run without breaking the vacuum. The thickness of each layer is controlled by
opening and closing the shutters in front of the sputter guns. The
base pressure of the system was lower than \SI{0.5}{\micro\pascal} with deposition 
pressures of \SI{1}{\pascal} for the Ta layers and
\SI{0.8}{\pascal} for the Co and Pt using Ar gas.

The seedlayers for the multilayer samples consist of
\SI{5}{\nano\meter} Ta and \SI{25}{\nano\meter} Pt. A bilayer of
\SI{0.3}{\nano\meter} Co and \SI{0.3}{\nano\meter} Pt is deposited
with 34 repetitions resulting in a \SI{20(1)}{\nano\meter}
\footnote{the value inbetween brackets is the uncertainty on the
  measured value in terms of the last digit, so
  \SI{20(1)}{nm}=20$\pm$\SI{1}{nm}. Similarly
  \SI{75.08(2)}{kA/m}=75.08$\pm$\SI{0.02}{kA/m}} magnetic layer. The
capping for the samples consists of \SI{3}{\nano\meter} Pt, which
prevents oxidation of the Co.

\subsection{Patterning of arrays of islands}

Laser interference lithography (LIL) is used to create a pattern 
in a photoresist layer, which acts as an etching mask.  

The pattern is first transferred into the bottom anti-reflective
coating (BARC) by O$_2$ reactive ion beam etching (RIBE). The BARC 
layer (DUV-30 J8) improves the resist pattern by limiting standing waves caused by interference of the incoming
waves with reflections from the metal layers. The pattern is then transferred into the magnetic
layer by Ar ion beam etching (IBE). All etching steps were performed in an
Oxford i300 reactive ion beam etcher.

After etching, the resulting samples have a Ta/Pt seedlayer
with magnetic islands on top. The average diameter of the
islands is approximately \SI{220}{\nano\meter} with a centre-to-centre
pitch of \SI{600}{\nano\meter}.

A lithography process is used to define Hall cross structures in a
layer of photoresist, similar as in our previous
work\cite{Engelen2010ahe}. The Hall cross structures are transferred
into the insulating layer using Ar IBE to ensure that during the Hall
measurement, the current only runs through a small ensemble of
islands.

The resulting structure consists of a conducting Hall cross of Ta/Pt with
magnetic islands with a diameter of \SI{220}{\nano\meter} and a pitch
of \SI{600}{\nano\meter} on top as shown in the SEM micrograph
in Figure~\ref{fig:HallCross}.

\subsection{Temperature dependent AHE}

The anomalous Hall measurements are performed in an Oxford
superconducting magnet. Using a temperature controller and a cryostat,
the measurements are taken between \SI{5}{\kelvin} and
\SI{300}{\kelvin}.  The magnetic field is applied perpendicularly to
the sample plane. An AC current at a frequency of \SI{12333}{\hertz}
is applied to the Hall cross, and the Hall signal is measured using a
lock-in amplifier.

For the statistical measurements of Figure~\ref{fig:TdepStatPlot}, the
switching field is measured over 150 times. During the acquisition,
the temperature variation from the setpoint is less than
\SI{0.1}{K}. The measurements are performed with a field sweep rate
$R$ of \SI{39}{\ampere\per\meter\per\second} at \SI{300}{K} and
\SI{3.9}{\ampere\per\meter\per\second} at \SI{10}{K}.

Since the variation of the switching field with temperature differs
between islands, the order of switching can change with the
temperature. This is especially true for weak islands. We took great
care to avoid mix-ups by comparing the step heights in the hysteresis
loops, so that the island we measure at \SI{10}{K} is the same island
as the one we measure at \SI{300}{K}. Since the mechanism which causes
the weak and strong island to differ is expected to be the same for
each island, an accidental mix up between two islands with similar
switching field will have limited effect on the final results. The
switching fields of a strong island are separated further apart, and a
change in reversal order is unlikely.

For the temperature dependent measurement, shown in
Figure~\ref{fig:TdepAHEPlot}, the field is swept between sample
saturation levels at a constant rate $R$ of
\SI{39}{\ampere\per\meter\per\second}. The temperature is kept
constant during the measurement and deviations from the setpoint are
less than \SI{0.5}{K} at the switching event.

\subsection{Magnetic characterisation}
The temperature dependence of the saturation magnetisation of the
continous, unpatterned film ($M_\text{s}(T)$) is determined using a
VSM. The sample temperature is regulated using a flow of nitrogen
cooling gas and a heater element.

The effective anisotropy at room temperature is determined by a home
built torque magnetometer. A DMS VSM-10 is used to determine the
temperature dependence of the effective anisotropy from the saturation
field, using the torque measurement at room temperature as scaling
factor.

\section{Results}

\subsection{Temperature dependent reversal}

\begin{figure}
  \centering
  \includegraphics[width=\widefigurewidth] {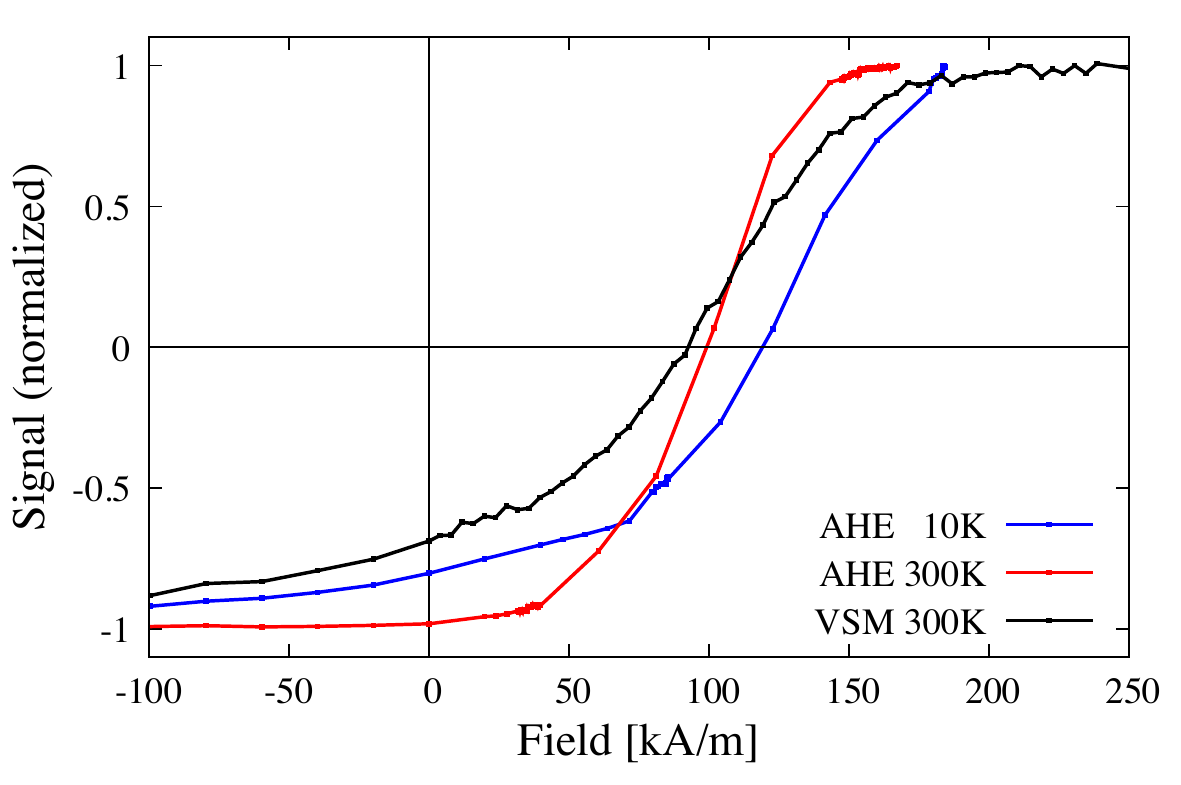}
  \includegraphics[width=0.45\widefigurewidth]{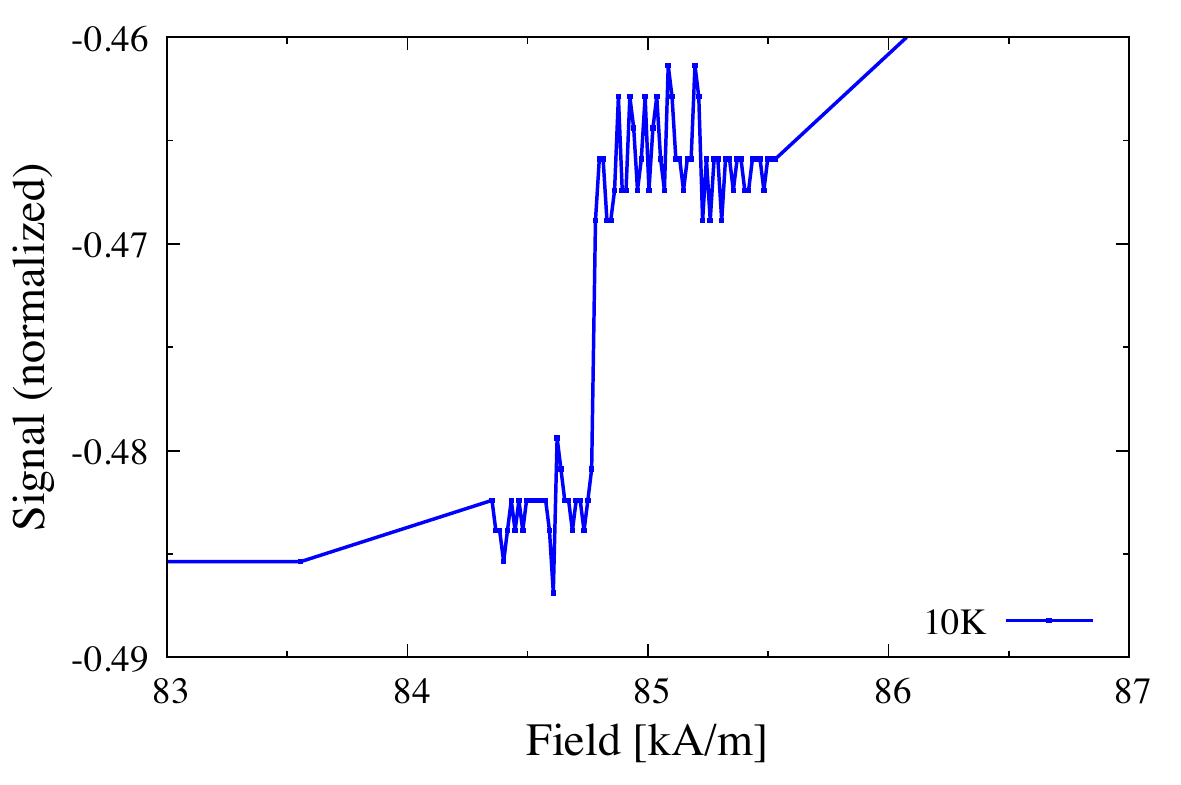}
  \includegraphics[width=0.45\widefigurewidth] {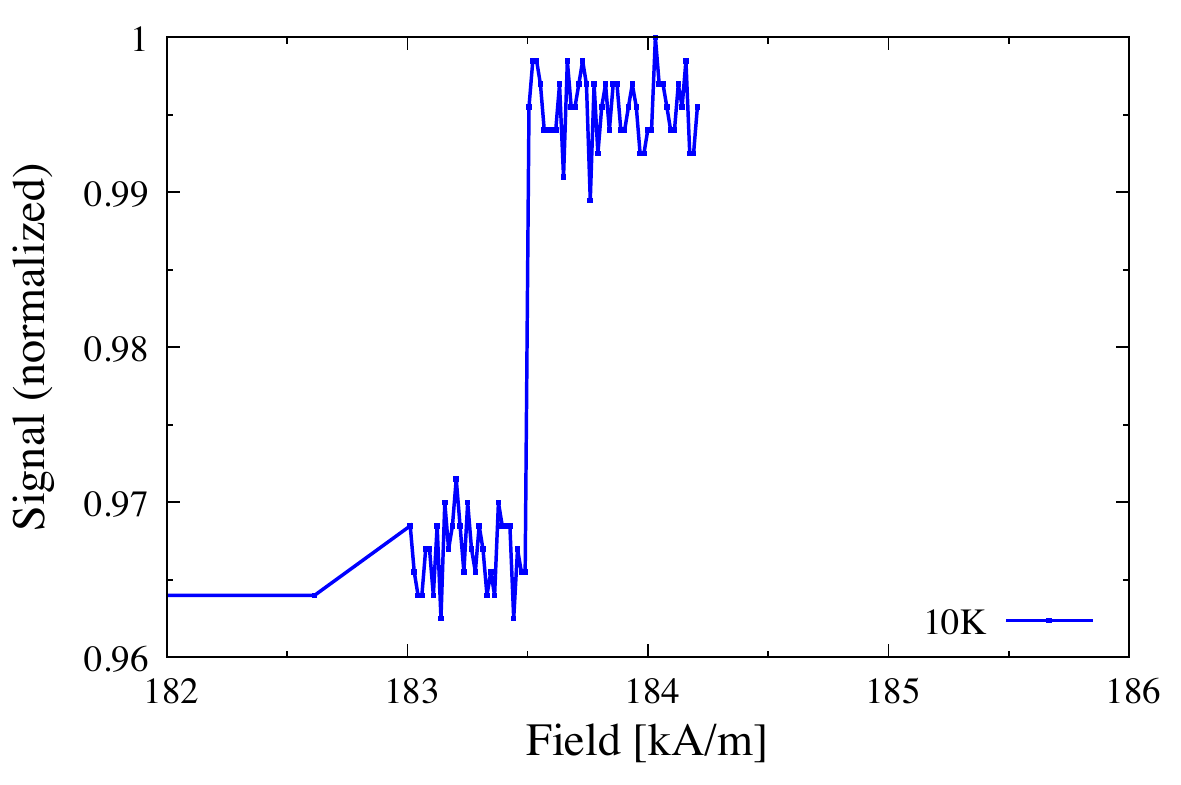}
  \includegraphics[width=0.45\widefigurewidth]{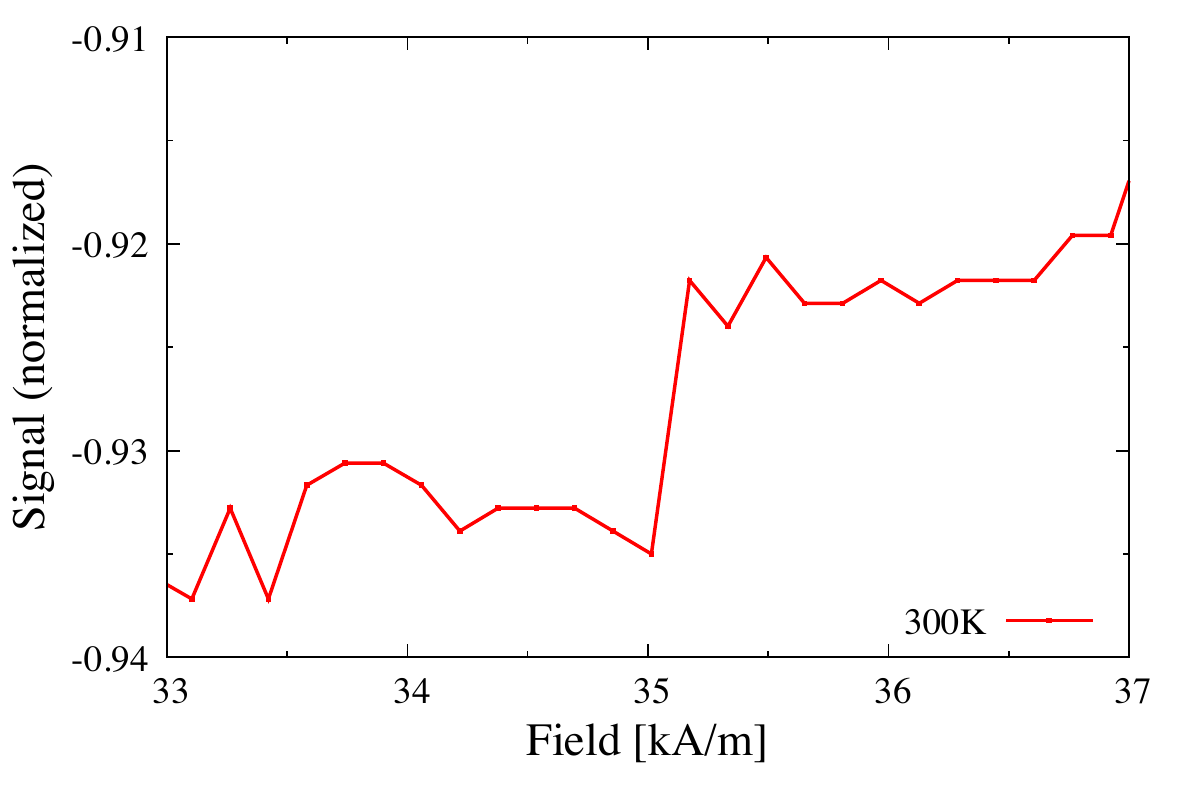}
  \includegraphics[width=0.45\widefigurewidth]{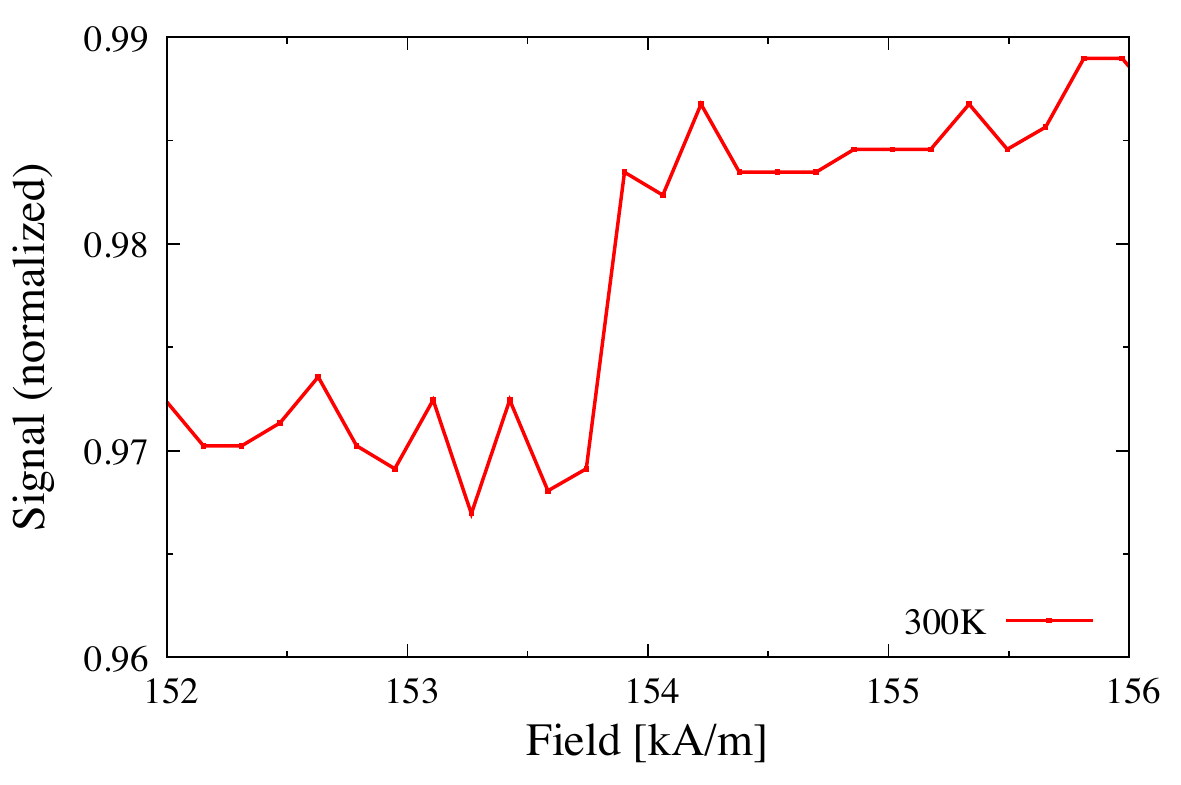}
  \caption{Top: Upward branch of AHE hysteresis curves at room temperature and
    \SI{10}{K} of an array of \num{80} islands, compared to a VSM
    hysteresis curve of a \num{8}$\times$\SI{8}{mm} sample with almost
    \num{200} million islands at room temperature. Bottom: In the AHE
    measurements, switching of individual islands can be observed. We
    compared weak islands, that switch at low fields, to strong
    islands switching at high fields.}
  \label{fig:VSMvsAHE}
\end{figure}

Figure~\ref{fig:VSMvsAHE} shows the upgoing part of the hysteresis
loop taken by AHE on the array of approximately \num{80} islands at
room temperature as well as \SI{10}{K}).  When the temperature is
decreased, the switching field increases. The AHE
measurements are compared to VSM measurements at room temperature of an
\num{8}$\times$\SI{8}{mm} sample with almost \num{200} million
islands. To enable comparison, the loops were scaled to the saturation
moment. The switching field distribution in the VSM loop
is higher, which can be attributed to the larger measurement area.

The AHE hysteresis loop shows small steps, which are caused by the
reversal of individual islands. The the field ramp rate is adapted in
such a way that we capture a switching event of a weak island,
switching at low field, and a strong island with a high switching
field. To save time, the intermediate field range is traversed more
quickly. Figure~\ref{fig:VSMvsAHE} shows four zooms, for a weak and a
strong island at 10 and \SI{300}{K}.

\subsection{Temperature dependence of the thermal switching field distribution}

\begin{figure}	
  \subfigure
  {\includegraphics[width=\widefigurewidth]
    {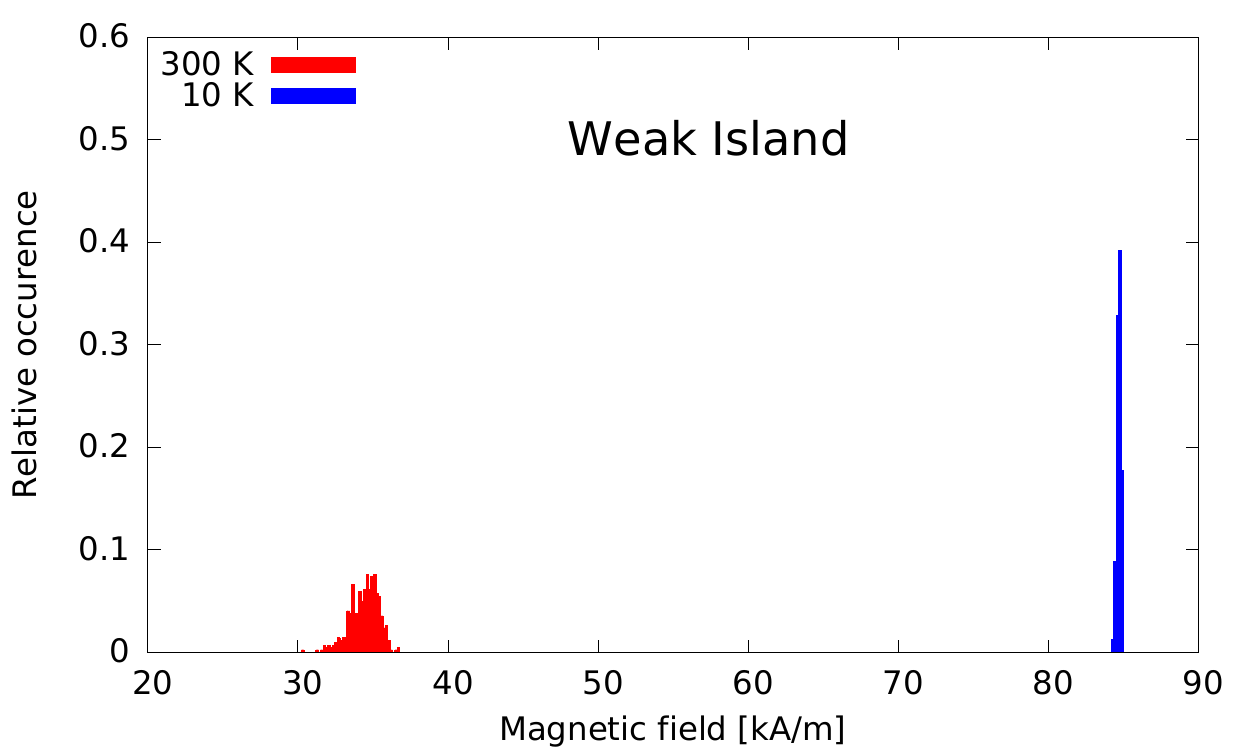}
    \label{fig:300K}
  }
  \subfigure  
  {
    \includegraphics[width=\widefigurewidth]
    {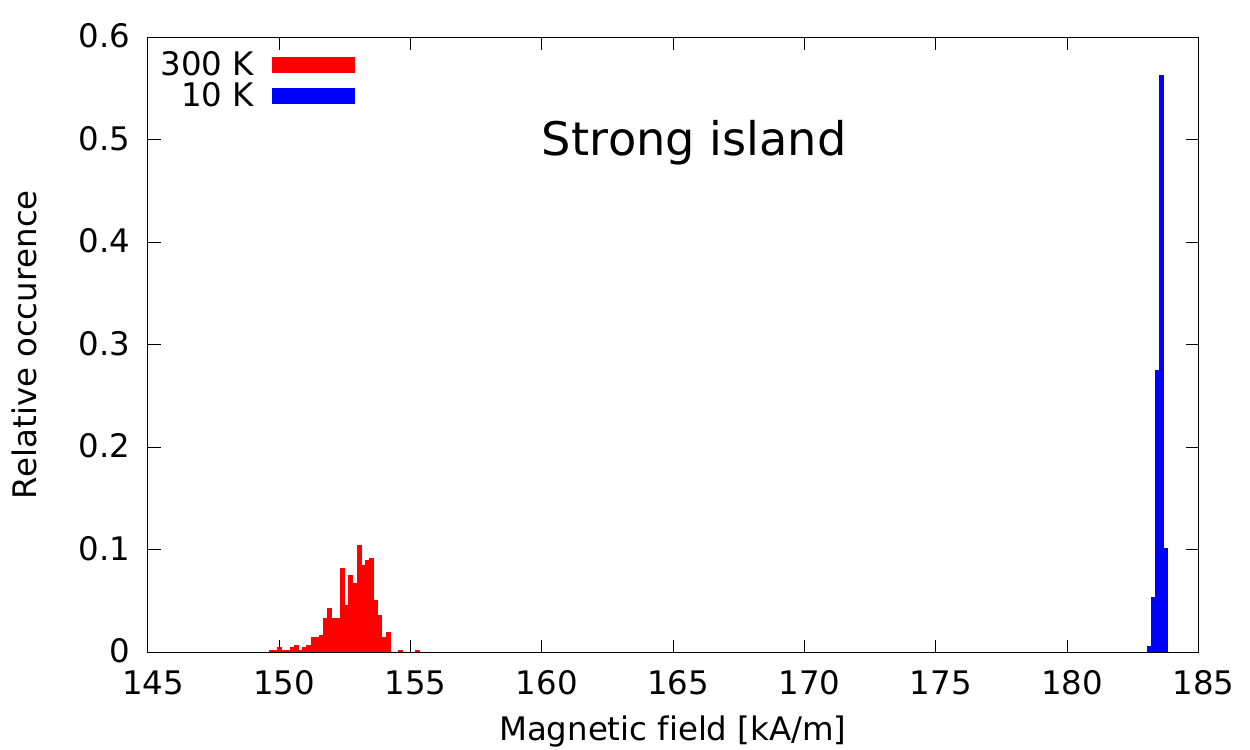}
    \label{fig:10K}
  }
  \caption{Histograms of the switching field of over 150 reversal
    incidents of a weak and strong island measured at 10.0(1)~K and
    300.0(1)~K. The bins are normalised to the total amount of
    reversals, so that the integral under the curves equals one. The
    width of each bin in the histogram is \SI{0.16}{\kilo\ampere\per\meter}.}
    \label{fig:TdepStatPlot}
\end{figure}

Figure~\ref{fig:TdepStatPlot} shows the thermally activated switching
field distribution at \SI{10}{K} and \SI{300}{K} for one of the first
islands that switches (weak) and one of the last islands (strong) when
ramping the field from -$H_\text{sat}$ to $H_\text{sat}$. During cooling,
two effects occur. In the first place, the average switching field
increases. Secondly, the width of the distribution decreases
dramatically. We can quantify this by dividing the full width at
half maximum of the distributions ($\Delta H$) by the field at
which the maximum in the distribution occurs ($H_\text{M}$). These
values are tabulated in Table~\ref{tab:FWHM} for the measurements on both
the strong and weak island. The relative distribution width
$\Delta H/H_\text{M}$ drops by one order of magnitude when the
temperature is decreased to \SI{10}{\kelvin}, which illustrates that
the origin of the variation in switching fields is indeed thermal
fluctuation. The same observation has been made for \SI{75}{nm} diameter
Co/Ni multilayered islands~\cite{Gopman2013}. 

\begin{table}
\begin{center} 
  \caption{Values for the full width at half maximum ($\Delta H$)
    divided by the field with the highest occurence ($H_\text{M})$) as
    a measure for the thermal switching field distribution.}
  \label{tab:FWHM}
  \begin{ruledtabular}
  \begin{tabular}{llcc}  \toprule
   & &{\SI{10}{\kelvin}} & {\SI{300}{\kelvin}}\\
  \\
    Weak & $\Delta H$ [\si{kA/m}]& 0.29 & 1.97\\
         & $H_\text{M}$ [\si{kA/m}] & 84.7 & 34.7\\
    	   & $\Delta H$/$H_M$ & 0.0034 & 0.057\\
        \midrule
    Strong & $\Delta H$ [\si{kA/m}]& 0.23 & 1.60\\
         & $H_\text{M}$ [\si{kA/m}]& 184 & 153\\
    	   & $\Delta H$/$H_M$ & 0.0012 & 0.010\\
 	\bottomrule
 	\end{tabular}
        \end{ruledtabular}
 	\end{center}
\end{table}

The distributions are fitted to Equation~\ref{eq:7}, with
$\Delta U_0$ and $H_\text{n}^0$ as fitting parameters. The
results of the fit are given in Table~\ref{tab:BigTable} under the
caption ``Statistical fit''. When decreasing the temperature from 300
to \SI{10}{K}, the switching field in the absence of thermal
activation $H_\text{n}^0$ increases. The increase is
more substantial for the weak island (40\%) than for the strong island
(9\%). The observed increase in the average switching field in
Figure~\ref{fig:TdepStatPlot} is therefore not only caused by a
reduction of thermal energy, other effects must also be taking place.

For both weak and strong islands, the energy barrier $\Delta U_0$
decreases upon cooling (by a factor of 2.7 and 3.7 respectively). From the values of
$\Delta U_0$ and $H_\text{n}^0$, we can calculate the domain wall
energy $\sigma_0$ and the width of the region of reduced domain wall
energy $w$, using the diamond model for reversal
(Equations~\ref{eq:18} and~\ref{eq:18b}). The substantial decrease in
the energy barrier seems to be caused by a strong decrease in $w$ (a factor
of two), with its ensuing decrease in the switching volume ($w^2t$), and, but much less so,
by a decrease in the domain wall energy (20\% to 50\% respectively).

\begin{table} \begin{center} \caption{Switching field H$_\text{n}^0$
      and energy barrier $\Delta U_0$ in the absence of thermal
      fluctuations determined from the fits to the thermal dependence
      of the switching field (in Figure~\ref{fig:TdepAHEPlot}) and from
      fitting the statistical measurements of the reversal of a weak
      and a strong island at \SI{10}{\kelvin} and
      \SI{300}{\kelvin} (Figure~\ref{fig:TdepStatPlot}).  From
      these values we can estimate the domain wall energy $\sigma_0$
      and the width of the reduced domain wall energy region $w$
      (Figure~\ref{fig:linearsigma}). The values in parentheses show
      the \SI{95}{\percent} confidence intervals obtained from the
      fit (H$_\text{n}^0$ and $\Delta U_0$) and combined measurement
      parameter errors ($w$ and $\sigma_0$).}
  \label{tab:BigTable}
  \begin{ruledtabular}
  \begin{tabular}{lcccc}
  \toprule
        & \multicolumn{2}{c}{Temperature fit} & \multicolumn{2}{c}{Statistical fit}\\
  	\\
        Weak & I & II & 10 K & 300 K\\
        \midrule
	H$_\text{n}^0$	[\si{kA/m}] & 
                                      75.08(2)& 93.14(1)& 87.28(3)&53.6(1)\\
 	$\Delta U_0$ [\si{\electronvolt}] &
                                       1.28(1) & 2.0(7) & 0.65(1) & 1.74(1))\\
 	$w$  [\si{nm}] & 7.9(3) & 9(2) & 5.2(3)&11.0(5) \\ 	
        $\sigma_0$ [\si{mJ/m^2}]& 0.65(1) & 0.9(2) &0.50(2)&0.64(4)\\
       \\
        Strong &   & I & 10 K & 300 K\\
        H$_\text{n}^0$	[\si{kA/m}] & 
                                      & 183(2)& 185.52(2)& 168.24(8)\\
 	$\Delta U_0$ [\si{\electronvolt}] &
                                       & 2.7(4) & 1.78(1) & 6.74(3))\\
        $w$  [\si{nm}]         & & 7.4(9) & 6.0(3)& 12.2(5) \\ 	
        $\sigma_0$ [\si{mJ/m^2}]& & 1.5(2) &1.20(5)&2.22(9)\\
 	\bottomrule
 	\end{tabular}
        \end{ruledtabular}
 	\end{center}
\end{table}

\subsection{Temperature dependence of average switching field}
\begin{figure}
  \centering
  \includegraphics[width=\widefigurewidth]
  {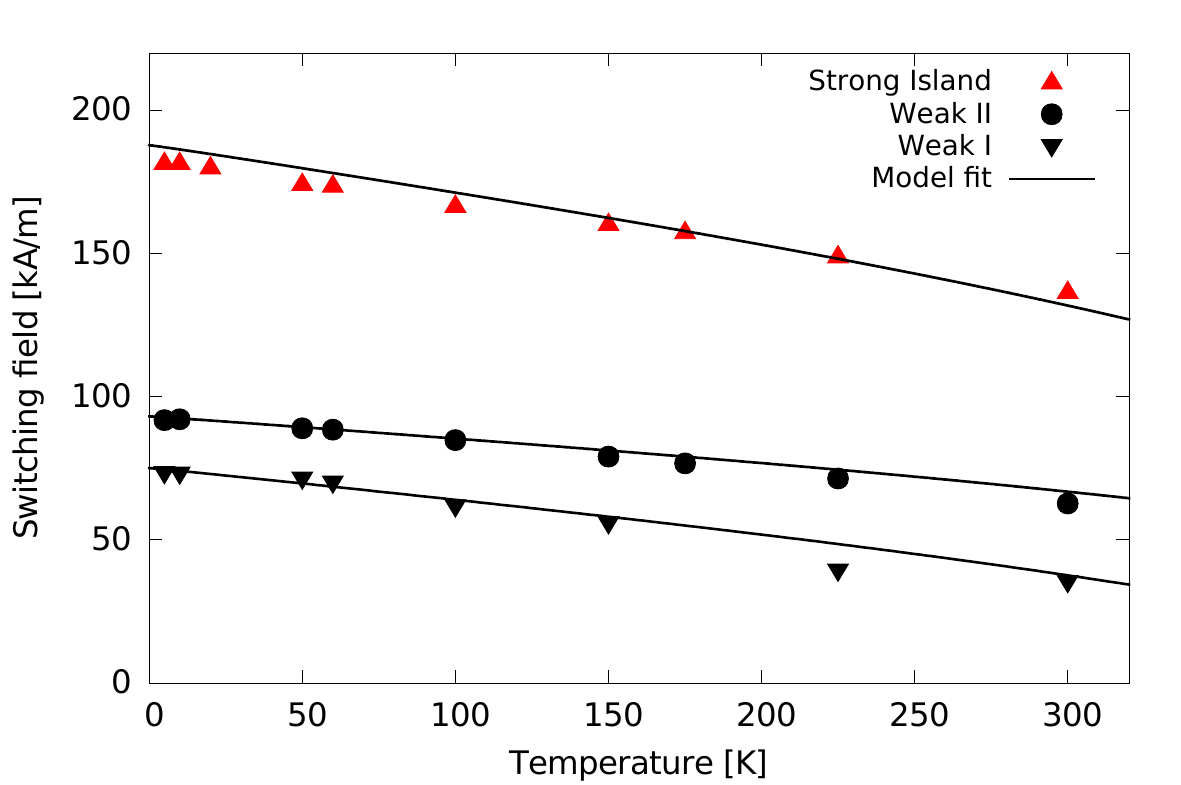}
  \caption{Temperature dependent average switching field for a weak and a strong
    island using temperature dependent AHE measurements. The lines are 
    fitted using Equation~\ref{eq:7} with the energy barriers given by
    Equation~\ref{eq:15} , under the condition that the
    switching volume $V_\text{sw}$ (Equation~\ref{eq:19}) is independent of temperature.
    \label{fig:TdepAHEPlot}}
\end{figure}

In addition to distributions at 10 and \SI{300}{K}, we used the anomalous
Hall effect to estimate the average switching field from single
hystersis loops. Figure~\ref{fig:TdepAHEPlot} shows the temperature
dependence of the average switching field of one strong and two weak
islands. The measurements are fitted to the theory from
Equation~\ref{eq:7}, using the energy barriers for the diamond model
(Equation~\ref{eq:15}) and under the restriction that the fitting
parameters do not change with temperature. 
  The figure shows that the actual
temperature dependence of the strong island is slightly lower than
predicted by the model, whereas that of the weak islands is slightly
higher. This is an indication that assuming the temperature independence
of the material parameters is incorrect.

The fitting parameters $H_n^0$ and $\Delta U_0$ are tabulated in
Table~\ref{tab:BigTable} under the caption ``Temperature Fit''. The
values of $H_\text{n}^0$ agree well with those obtained from the
distributions at \SI{10}{K}, but are higher than those obtained at
\SI{300}{K}. The energy barrier estimated from the temperature
dependence of the average switching field is, however, much larger
than that obtained from the distribution at \SI{10}{K}. This again
clearly demonstrates that assuming temperature independent material
parameters leads to incorrect conclusions about the thermal stability
of the islands. The value of the energy barrier of the strongest
island is in agreement with that estimated by Kikuchi \emph{et
  al}~\cite{Kikuchi2011} (\SI{5.5}{eV}) on a \SI{300}{nm} diameter
island prepared from a [Co(\SI{0.9}{nm})/Pt(\SI{2}{nm})]$_3$
multilayer. The estimate for the nucleation field in the absence of
thermal fluctuations is \SI{0.4}{MA/m}, which is higher by a factor of
two than that in our experiment. The difference could be caused by the
better defined interfaces, due to the thicker Co layer, and the
reduced number of bilayers. One should however also take into
consideration that in their work, a coherent rotation model was
assumed, which leads to higher values for the estimate of the energy
barrier and the switching field~\cite{Engelen2010ahe}.

\subsection{Temperature dependence of the material parameters}

To gain insight into the temperature dependence of the material
parameters, we measured  VSM hysteresis loops from \SI{170}{K} up to
room temperature. From these loops the saturation magnetisation and
anisotropy are estimated. 

\subsubsection{Temperature dependence of the saturation magnetisation}
\begin{figure}
    \centering
    \includegraphics[width=\widefigurewidth]{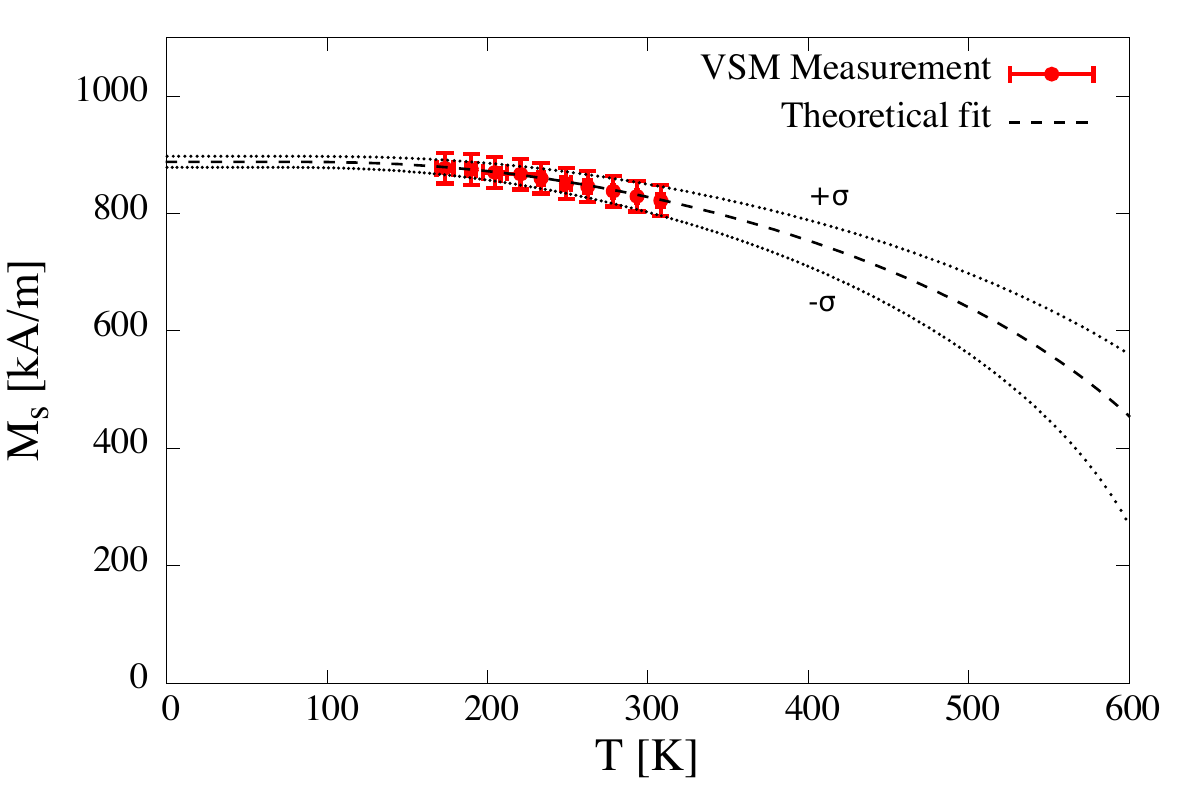}
    \caption{Temperature dependence of $M_\text{s}$ 
    from temperature dependent VSM measurements on the continuous film and 
    a fit using Equations~\ref{eq:1}~and~\ref{eq:2}. The $\sigma$
    lines indicate the 68.2\% confidence intervals.}
    \label{fig:MsPlot}
\end{figure}

Figure~\ref{fig:MsPlot} shows that indeed the saturation magnetisation
decreases slightly with increasing temperature. The curve is fitted
to the Brillouin function (Equations~\ref{eq:1} and \ref{eq:2}), with
fitting parameters the Curie temperature ($T_\text{c}$) and the
saturation magnetisation at 0~K ($M_\text{s}(0)$). The Curie
temperature is estimated to be \SI{684(58)}{K}, which is in agreement
with previous studies of Co/Pt multilayers~\cite{Meng1996,
  Kesteren1993}. The value of $M_\text{s}(0)$ is estimated to be
\SI{888(9)}{kA/m}. To estimate the errors of the fit, a Monte Carlo
method is used, where we assumed $\sigma_{T}$=\SI{7}{\kelvin} and
$\sigma_{M_\text{s}(T)}$=\SI{10}{\kilo\ampere\per\meter}.

From the fit we can conclude that the saturation magnetisation
decreases by about 7\% when increasing the temperature from 10 to
\SI{300}{K}. By itself, this is not sufficient to explain the large
variation in $H_\text{n}^0$. Shan~\emph{et al.}~\cite{Shan1994} report a much
stronger decrease in magnetisation, by 22\%, for a similar Co layer
thickness, but much thicker Pt thickness (\SI{1.5}{nm}). The dependence
they measured however does not resemble a Brillouin function.

It any case, it is clear that the magnetisation changes, and we may expect that
 other material parameters change as well. Therefore, we also estimated the
anisotropy from the VSM hysteresis loops.

\subsubsection{Temperature dependence of the anisotropy}

\begin{figure}
  \centering
  \includegraphics[width=\widefigurewidth]
  {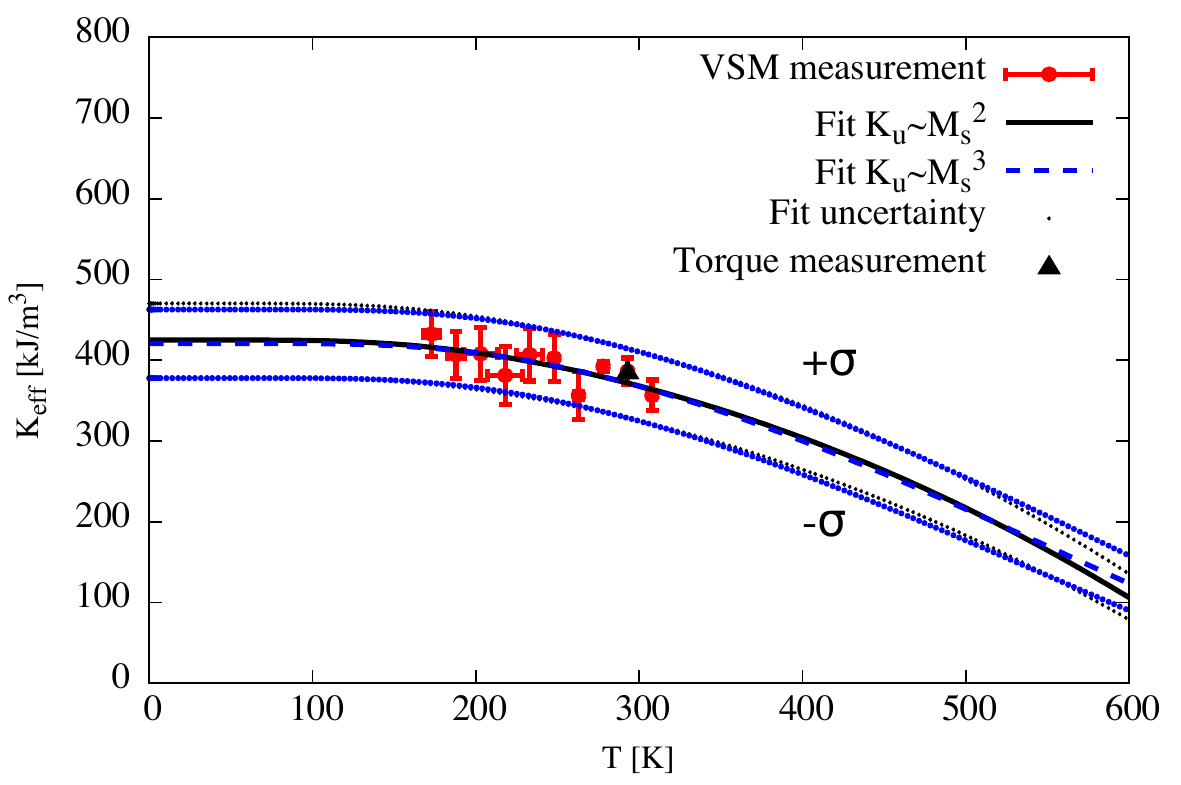}
  \caption{Temperature dependent anisotropy VSM measurements and
    torque measurement at room temperature on the continuous film. The
    theoretical fits are created using
    Equation~\ref{eq:Anis} for exponent $n$=2 or 3. The $\sigma$ lines
    indicate the 68.2\% confidence intervals of the fits.}
  \label{fig:AnisotropyPlot}
\end{figure}

Figure~\ref{fig:AnisotropyPlot} shows the anisotropy of the continuous
film as a function of the temperature, obtained by VSM, using a room
temperature torque measurement for
calibration. Equation~\ref{eq:Anis} is fitted to the measured
$K_\text{eff}(T)$ with the fitted parameters $T_\text{c}$,
$M_\text{s}(0)$ and $K_\text{u}(0)$, where the value of the exponent
$n$ is either of the two extremes. Given the uncertainty in the estimate of the anisotropy,
and the limited temperature range, it is impossible to
determine which exponent is
correct. Table~\ref{tab:AnisotropyFitTable} shows the fitted parameters
for both cases. To obtain an estimate of the measurement errors in the
fitted parameters, again a Monte Carlo method is used, for which we
assumed the uncertainty in the temperature to have a standard deviation of
\SI{3.5}{K}, and \SI{6}{kJ/m^3} in the values of $K_\text{eff}$.

\begin{table} 
  \begin{center} 
    \caption{Fitted parameters for the temperature dependence of the
      effective anisotropy (Figure~\ref{fig:AnisotropyPlot}), assuming
      the effective anisotropy is proportional to $M_\text{s}^2$ or $M_\text{s}^3$.}
  \label{tab:AnisotropyFitTable}
  \begin{ruledtabular}
  \begin{tabular}{lcc}
  \toprule
        &$n$=2 & $n$=3\\
  	\midrule
	$K_\text{u}$ [\si{MJ/m^3}] & \SI{0.91(2)}{}& \SI{0.89(2)}{}\\
        $M_\text{s}$ [\si{MA/m}] & \SI{0.89(2)}{}& \SI{0.86(2)}{}\\
        $T_\text{c}$ [\si{K}] & \SI{679(15)}{}& \SI{872(20)}{}\\
 	\bottomrule
 	\end{tabular}
        \end{ruledtabular}
 	\end{center}
\end{table}

The exact value of the exponent in Equation~\ref{eq:Anis} has very
little effect on the estimate of the anisotropy and magnetisation of
the film. Assuming the exponent to lie somewhere between the two
extrema, the fitted value of $M_\text{s}(0)$=\SI{0.87(4)}{MA/m}
is equal, within the estimation error, to the value found from
extrapolation of the $M_\text{s}(T)$ curve (\SI{0.888(9)}{MA/m}, see
Figure~\ref{fig:MsPlot}). The similarly estimated value of $K_\text{u}(0)$ is
\SI{0.90(3)}{MJm^{-3}}. 

The value of the exponent does however have a significant effect on
the estimate for the Curie temperature $T_\text{c}$. The estimated
Curie temperature for $n$=2 is, within the measurement error, equal to
the estimate we obtained from the temperature dependence of the
magnetisation (\SI{684(58)}{K}), which is in agreement with previous
studies. This suggests that the intrinsic anisotropy in the film
($K_\text{u}$) is rather more proportional to $M_\text{s}^2$ than to
$M_\text{s}^3$. This seems to indicate that the origin of the
perpendicular anisotropy is not only due to interface anisotropy.  A
wider temperature range might help to narrow down the estimates, but
it should be noted that at temperatures above \SI{500}{K}, the Co/Pt
interfaces start to mix.

\section{Discussion}
When we apply the diamond model to the measured thermally activated
switching field distributions, we conclude that difference between
islands is primarily caused by a difference in wall energy $\sigma_0$,
and much less due to a difference in transition region $w$. Why the
domain wall energy varies between islands cannot be determined from
Anomalous Hall measurements only. A possible cause for a reduction in
wall energy might be edge damage caused by the ion beam etching
process, leading to mixing between the Co and Pt layers and loss of
interface anisotropy. Also edge roughness caused by the lithography
might play a role, since it will strongly influence the way the domain
wall enters the island.

Based on previous reports and our observations, we can conclude that
both the magnetisation and anisotropy decrease with increasing
temperature. It is very unlikely that the exchange constant $A$
increases, since it generally decreases with
magnetisation~\cite{Goll2004}. Since the wall energy is proportional
to $\sqrt{A K_\text{u}}$, it should decrease as the temperature
increases. This is in contradiction with the fit to the distributions,
from which we conclude that the wall energy \emph{increases} by 20\%
to 50\%. The origin of this apparent discrepancy should be the subject
of further study.

In addition to a moderate increase in the domain wall energy, our
simple diamond model predicts a strong increase of the switching
volume with increasing temperature.  If the region of reduced wall
energy $w$ is somehow related to the wall thickness (proportional to
$\sqrt{\nicefrac{A}{K_\text{u}}}$), we would also expect a large
variation in the wall energy (proportional to
$\sqrt{AK_\text{u}}$). However, this is not the case. If $w$ is due to
etch damage during the fabrication process or edge roughness, there is
a possibility that the temperature dependency remains, or even
increases. This point also deserves further investigation.

The interpretation of the thermally activated distributions depends on
having good models for the thermal stability (Equation~\eqref{eq:7}) and the
relation between the energy barrier and the strength of the applied field
(Equation~\eqref{eq:15}). Since the model for thermal stability is
well established, and fits almost perfectly to the distributions, we
assume that it is correct. Our diamond model is simple, but a more
elaborate micromagnetic model, along the lines of
Adam~\cite{Adam2012}, will not resolve the above contradictions since
it is based on the same assumptions and differs only in a more
realistic island shape and anisotropy distribution. For furher
refinement, one might have to include the possibility that
reversal can take place over multiple pathways~\cite{Roy2017}, each of
which can have a different temperature dependence. 

It is without doubt, however, that the temperature dependence of the material
parameters is substantial. Determining the energy
barrier from the distribution of the switching fields of the individual
islands at the temperature of interest is therefore to be preferred.

\section{Conclusion}

By means of the very sensitive anomalous Hall effect, we have been
able to measure the reversal of individual magnetic islands of
diameter \SI{220}{nm} in an array of approximately eighty islands with a
centre-to-centre pitch of \SI{600}{nm}. By traversing the hysteresis loop more than 150
times, we have observed that the switching field of an individual
island fluctuates by about \SI{10}{kA/m}. When reducing the
temperature, this variation for a single island
decreases significantly, which proves that the cause of the fluctuations is
thermal energy in the system.

From the distribution in switching fields of a single island, we can
estimate the switching field in the absence of thermal fluctuations,
$H_n^0$, and the energy barrier at zero field, $\Delta U_0$. This
estimate requires a model that relates the decrease in the energy barrier
with an increase in the externally applied field. We developed a simple
``diamond'' model, based on creation and the subsequent propagation of a
domain wall. Reasonable nucleation fields can only be achieved if we
assume the domain wall energy to increase from zero as the wall
enters the island, up to a maximum value after a certain distance $w$
from the edge of the island.

From the model fit to the thermal switching field distributions, we
estimate that $H_n^0$ decreases when the temperature increases from
10 to \SI{300}{\kelvin}.  The temperature dependence is
more prominent for weak islands (approximately 40\%), than for strong
islands (approximately 10\%). The energy barrier on the other hand
has a much stronger dependence (it increases by a factor of three to
four). Translated to the parameters of the diamond model, the increase
in the energy barrier is mainly due to an increase in the switching volume
($w^2t$), which increases by a factor of two, and much less due to an
increase in the domain wall energy ($\sigma_0$), which increases by 20\% to 50\%.

The switching field in the absence of thermal energy, $H_n^0$, does
not necessarily have to be identical to the switching field measured at
\SI{0}{K}, since the material parameters will vary with temperature. When
we extrapolate the switching field to \SI{0}{K}, we find values that
are almost identical to $H_n^0$ measured at \SI{10}{K}, which is
substantially higher than the $H_n^0$ measured at \SI{300}{K}. The energy
barrier determined from the dependence of the switching
field on temperature is also strongly overestimated, by at least a
factor of two for the weakest island and 30\% for the strongest compared to the 
measurement at \SI{10}{K}.

That the material parameters do vary with temperature is illustrated by
temperature dependent VSM measurements, which show that the 
magnetisation decreases by \SI{7}{\percent} and anisotropy by \SI{16}{\percent} 
when increasing the temperature from 10 to \SI{300}{K}.

However, whichever method is used, the value of $w$ is similar for weak and
strong islands and varies from 5 to \SI{12}{nm}. The domain wall
energy for weak islands  (0.5 to \SI{1}{mJ/m^2}) is clearly lower than that for
strong islands (1.2 to \SI{2.3}{mJ/m^2}). Within the framework of our
model, we must therefore conclude that the variation in the switching
fields between islands must be caused by variations in
domain wall energy.

Our work demonstrates that detailed observations of the fluctuations in
the switching fields of individual islands allows us to determine the basic
parameters of the energy barrier between magnetisation states, such as
the height of the energy barrier (important for thermal stability) and
the field required to overcome this barrier in the absence of thermal
fluctuations (important for ultra-fast switching). In contrast to
temperature dependent measurements, which rely on the assumption
that the material parameters are temperature independent, our method
allows us to determine these parameters at any temperature. This is important, for
instance, for applications working at room temperature,
such as data storage in bit patterned media, magnetic random access
memories, and magnetic logic circuits.

\section*{Acknowledgments}
The authors wish to thank Henk van Wolferen and Johnny Sanderink for
fabrication support, Dr. N. Kikuchi of Tohoku University and
Prof. T. Thomson of Manchester University for valuable discussions and
the low temperature VSM measurement of figure~\ref{fig:VSMvsAHE},
and proof-reading-service.com for an exceptional job in manuscript
correction. This research was supported by the Dutch Technology
Foundation STW, which is part of the Netherlands Organisation for
Scientific Research (NWO), and which is partly funded by the Ministry
of Economic Affairs.

\bibliographystyle{bst/apsrev_modified_doi}
\bibliography{paperbase}

\end{document}